\documentclass[10pt]{article}
\usepackage{mathrsfs}
\usepackage{amsmath,amsfonts,amssymb,mathtools}
\usepackage{cite}
\usepackage{dsfont}
\usepackage{graphicx}
\usepackage{hyperref}
\usepackage{verbatim}
\usepackage{slashed}
\usepackage[all]{xy}
\usepackage{xcolor}
\usepackage{subfig}
\usepackage{tikz}
\usetikzlibrary{shapes.geometric,arrows}
\usepackage{relsize}
\usepackage{caption}
\usepackage{float}
\usetikzlibrary{positioning}
\usepackage{geometry}
\usepackage[T1]{fontenc}
\usepackage{lmodern}
\usepackage{stackengine} 
 
\usepackage[utf8]{inputenc}

\hypersetup{
  colorlinks,
  citecolor=violet,
  linkcolor=blue,
  urlcolor=blue}

\csname @addtoreset\endcsname{equation}{section}
\DeclareMathAlphabet\mathbfcal{OMS}{cmsy}{b}{n}

\newcommand{\beq}{\begin{equation}}
\newcommand{\eeq}{\end{equation}}
\newcommand{\bea}{\begin{eqnarray}}
\newcommand{\eea}{\end{eqnarray}}
\newcommand{\ba}{\begin{array}}
\newcommand{\ea}{\end{array}}
\newcommand{\bit}{\begin{itemize}}
\newcommand{\eit}{\end{itemize}}
\newcommand{\nn}{\nonumber}

\newcommand{\mezzo}{\frac{1}{2}}
\newcommand{\complesso}{{\ \hbox{{\rm I}\kern-.6em\hbox{\bf C}}}}
\newcommand{\reale}{{\hbox{{\rm I}\kern-.2em\hbox{\rm R}}}}
\newcommand{\uno}{ \,  \raisebox{+0.14em}{{\hbox{{\rm \scriptsize ]}} \raisebox{-0.2em}{\kern-.8em\hbox{1}}}} \, }  

\newcommand{\p}{\partial}

\renewcommand{\a}{\alpha}
\renewcommand{\b}{\beta}
\newcommand{\g}{\gamma}

\newcommand{\G}{\Gamma}
\renewcommand{\d}{\delta}
\newcommand{\D}{\Delta}
\newcommand{\e}{\epsilon}
\newcommand{\Er}{{\mathbfcal{E}}}

\renewcommand{\l}{\lambda}
\renewcommand{\L}{\Lambda}

\newcommand{\m}{\mu}

\newcommand{\n}{\nu}
\renewcommand{\r}{\rho}
\newcommand{\s}{\sigma}

\renewcommand{\S}{\Sigma}

\newcommand{\ve}{\varepsilon}

\newcommand{\om}{\omega}
\newcommand{\Om}{\Omega}

\begin{document}


\begin{titlepage}

\vspace{0.3cm}

\begin{flushright}
$LIFT$--4-2.23
\end{flushright}

\vspace{0.3cm}

\begin{center}
\renewcommand{\thefootnote}{\fnsymbol{footnote}}
\vskip 10mm  
{\Huge \bf Accelerating and Charged  
\vskip 7mm
  Type I Black Holes  }
\vskip 30mm
{\large {Marco Astorino$^{a}$\footnote{marco.astorino@gmail.com} 
}}\\

\renewcommand{\thefootnote}{\arabic{footnote}}
\setcounter{footnote}{0}
\vskip 8mm
\vspace{0.2 cm}
{\small \textit{$^{a}$Laboratorio Italiano di Fisica Teorica (LIFT),  \\
Via Archimede 20, I-20129 Milano, Italy}\\
} \vspace{0.2 cm}

\end{center}

\vspace{4.1 cm}

\begin{center}
{\bf Abstract}
\end{center}
{A new, exact and analytical class of accelerating and charged black holes is built, in the Einstein-Maxwell theory, thanks to the Harrison transformation. The diagonal metric does not belong to the Petrov type D classification, therefore it is not part of the Plebanski-Demianski spacetimes. The simplest subcase of this family recovers the Reissner-Nordstrom black hole in the vanishing  acceleration limit and  the standard C-metric in the limit of null electric charge. More general cases can have two independent electric charges, which can be tuned as desired, even to remain with an uncharged black hole, such as Petrov Type I Schwarzschild, embedded in an accelerating charged Rindler background. These accelerating black holes can be considered as a limit of charged binary systems. 
Conical singularities can be possibly removed in extremal configurations. \\
The entropy of the conformal field theory model dual to the extreme black hole is obtained from near horizon analysis. Magnetic, dyonic, NUTty and Kerr-like extensions are also discussed.
}

\end{titlepage}

\addtocounter{page}{1}

\newpage

\tableofcontents
\newpage

\section{Introduction}

Accelerating black holes usually are considered to fall into the Plebanski-Demianski family of solutions \cite{Plebanski-Demianski}, thus they belong to the type D in the Petrov classification\footnote{Accelerating black hole embedded in external gravitational \cite{ernst-generalized-c} or electromagnetic \cite{ernst-remove} backgrounds or multi black hole configurations \cite{multipolar-acc} were already know to be of more general type with respect to the D-type, however here we are referring only to accelerating metrics that models a black hole with a Rindler horizon.}. Only recently it has been shown that accelerating Taub-NUT black holes, first found in \cite{mann-stelea-chng}, are of type I \cite{Podolsky-nut}. Then in \cite{PD-NUTs} it has been shown that actually all the black holes of the Plebanski-Demianski class can be generalised to the type I and endowed with a NUT charge thanks to the Ehlers transformation. From a physical point of view it seems that the Ehlers map also affects the accelerating horizon embedding the initial black hole into a Rindler-NUT background. It's worth noting that a subclass of Plebanski-Demianski metrics can simultaneously have acceleration and NUT charge, these are the solutions with non-null angular momentum. However, as shown in  \cite{PD-NUTs},  NUTty solutions with angular momentum and acceleration generated by the Ehlers transformation are of a different kind with respect to the usual ones, indeed they differ even on the Petrov type. This fact does not happen without the accelerating horizon, for instance, the Harrison transformation maps the Kerr metric into the Kerr-NUT or maps the Kerr-NUT in itself.     \\
A transformation similar to the Ehlers for axisymmetric and stationary spacetimes is known, the Harrison transformation. Both transformations are non-trivial Lie-point symmetries of the Ernst equation, which give an alternative (and equivalent to the Einstein-Maxwell equations) description of fields for the theory of general relativity in the presence of two commuting Killing vectors. These symmetries of the Ernst equations allow one to generate basically all solutions of the theory by applying them to an initial given solution, which is denominated as seed.  The Ehlers transformation is known to rotate the seed mass into the gravitomagnetic mass \cite{reina-treves}, \cite{enhanced}, \cite{PD-NUTs}, basically adding the NUT charge to a massive solution. On the other hand the Harrison transformation is known to add the electromagnetic charge to the seed. It is well known that it can generate from the Schwarzschild seed the Reissner-Nordstrom black hole or to charge the double static black hole to get a charged Bach-Weyl solution \cite{many-rotating}.  \\
Because of the similarities between the Ehlers and the Harrison transformation and since the non-trivial and non-intuitive behaviour the former transformation has in the presence of an accelerating horizon a natural question arises: What is the effect of the Harrison transformation on accelerating metrics? What one naively expects is that the Harrison symmetry acting on the C-metric should generate the well known charged C-metric. Nevertheless if the analogy with the Ehlers transformation is strong we may encounter  some non-intuitive and novel results. This matter is discussed in detail in section \ref{sec:generation}, for a simple case in the presence of electrical charge only; also the possible removal of the conical singularity typically affecting the accelerating black holes is discussed. For the dyonic case see appendix \ref{app:dyonic-acc-RN-I}, which also contains some details on the nature of the Harrison transformation and a proposal for an enhanced version of it. In section \ref{sec:generation-NUT} we combine both the Ehlers and the Harrison maps to build more general solutions, looking for possible connections with the charged and NUTty type I black hole of \cite{PD-NUTs}.
We start, in the next section, with a very brief review of the Ernst equations and the Harrison transformation.\\

\section{Ernst equations and the Harrison transformation}
\label{sec:Ernst-Harrison}

\subsection{Electrovacuum Ernst Equations}
Consider general relativity coupled with Maxwell electromagnetic field in four space-time dimensions. The theory is governed by the Einstein-Hilbert action
\beq
         I[g_{\m\n},A_\m] = - \frac{1}{16\pi G} \int_\mathcal{M} d^4x \sqrt{-g} \left(R - F_{\m\n}F^{\m\n} \right) \ , \nn
\eeq
whose variation with respect to the metric $g_{\mu\nu}$ and electromagnetic vector potential, $A_\m$ gives the Einstein-Maxwell field equations
\bea  \label{field-eq-g}
                        &&   R_{\m\n} -   \frac{R}{2}  g_{\m\n} =   F_{\m\r}F_\n^{\ \r} - \frac{1}{4} g_{\m\n} F_{\r\s} F^{\r\s}  \quad ,   \\
       \label{field-eq-A}                  &&   \partial_\m \big( \sqrt{-g} F^{\m\n} \big) = 0  \ \quad .
 \eea
The electromagnetic potential $A_{\m}$ defines, as usual, the Faraday tensor such that $F_{\m\n} = \p_\m A_\n -\p_\n A_\m$. We focus on gauge fields which preserve the axisymmetric and stationary symmetry of the metric, therefore we pick the vector  potential as follows
\beq \label{Am}
A_{\mu}=\big[ A_t (\r,z) , \  0 , \ 0 , \ A_\varphi (\r,z) \big] \ \ .
\eeq
For the theory under consideration, the most general four-dimensional axisymmetric and stationary spacetime, which thus possesses two commuting Killing vector fields $(\p_t,\p_\varphi)$, is  the Lewis-Weyl-Papapetrou metric
\beq
\label{LWP-metric}
{ds}^2 = -f ( dt - \omega d\varphi)^2 + f^{-1} \bigl[ e^{2\gamma}  \bigl( {d \rho}^2 + {d z}^2 \bigr) +\rho^2 d\varphi^2 \bigr] \ \ .
\eeq
All the functions of the metric $f,\omega,\g$, of the electric and magnetic potentials $A_t , A_\varphi$  depend only on the non-Killing coordinates ($\rho,z$), in order to respect the above symmetry requirements. Ernst has shown in \cite{ernst2} that Einstein-Maxwell field equations (\ref{field-eq-g})-(\ref{field-eq-A}) are essentially equivalent to the complex equations
 \bea
     \label{ee-ernst-ch}  \left( \textsf{Re} \ \Er + | \mathbf{\Phi} |^2 \right) \nabla^2 \Er   &=&   \left( \overrightarrow{\nabla} \Er + 2 \ \mathbf{\Phi^*} \overrightarrow{\nabla} \mathbf{\Phi} \right) \cdot \overrightarrow{\nabla} \Er   \quad ,       \\
     \label{em-ernst}   \left( \textsf{Re} \ \Er + | \mathbf{\Phi} |^2 \right) \nabla^2 \mathbf{\Phi}  &=& \left( \overrightarrow{\nabla} \Er + 2 \ \mathbf{\Phi^*} \overrightarrow{\nabla} \mathbf{\Phi} \right) \cdot \overrightarrow{\nabla} \mathbf{\Phi} \quad .
\eea
The electromagnetic and gravitational Ernst potentials, related to the metric (\ref{LWP-metric}) and electromagnetic potential (\ref{Am}), are complex functions of $\r$ and $z$, respectively defined by
\beq \label{def-Phi-Er}
       \mathbf{\Phi} := A_t + i \tilde{A}_\varphi  \qquad , \qquad \qquad     \Er := f - \mathbf{\Phi} \mathbf{\Phi}^* + i h  \quad ,
\eeq
where $\tilde{A}_t(\r,z)$ and $h(\r,z)$ stem from the following definitions\footnote{The following ordering ($\overrightarrow{e}_\r , \overrightarrow{e}_\varphi , \overrightarrow{e}_z$) of the three-dimensional basis has been chosen.}
\bea
    \label{A-tilde-e} \overrightarrow{\nabla} \tilde{A}_\varphi &:=&  \frac{f}{\r} \overrightarrow{e}_\varphi \times (\overrightarrow{\nabla} A_\varphi + \omega  \overrightarrow{\nabla} A_t ) \ \ , \\
    \label{h-e}    \overrightarrow{\nabla} h &:=& - \frac{f^2}{\r} \overrightarrow{e}_\varphi \times \overrightarrow{\nabla} \omega - 2 \ \textsf{Im} (\mathbf{\Phi}^*\overrightarrow{\nabla} \mathbf{\Phi} )  \ \ .
\eea
In principle an extra couple of differential equations should be taken into account to determine $\g(\r,z)$. These are not included in the Ernst equations (\ref{field-eq-g})-(\ref{field-eq-A}), because they are completely decoupled from the main system, so they can be subsequently considered after the functions $f,\omega,A_t,A_\varphi$ are established. Then these equations reduce to a pair of integrals, which can be solved to obtain the remaining unknown $\g(\r,z)$, see \cite{enhanced} for details. In any case, in the case under consideration in this article the $\g$ function remains fixed, thus it has not to be found. More generally it can be proven that under a continuous symmetry transformation of the Ernst system also the integral equations for $\g$ remain invariant, so also $\g$ does not change. \\

\subsection{Harrison transformation}

The Ernst equations (\ref{ee-ernst-ch})-(\ref{em-ernst}) enjoy the invariance of the $SO(2,1)$ group of Lie-point symmetries. The symmetry transformation we will mainly focus in this article is the Harrison map
\beq \label{harrison}
\Er \longrightarrow \bar{\Er} = \frac{\Er}{1-2\a^*\mathbf{\Phi} -\a\a^* \Er} \qquad \quad ,  \quad  \qquad  \mathbf{\Phi} \longrightarrow  \bar{\mathbf{\Phi}} = \frac{\mathbf{\Phi} + \a \Er}{1-2\a^*\mathbf{\Phi} -\a\a^* \Er} \quad .
\eeq
It is easy to check that this transformation leaves the Ernst equations invariant.
The fundamental property of symmetry transformations of Ernst equations is that they are able to bring a given solution of the Einstein and Maxwell equations (\ref{field-eq-g})-(\ref{field-eq-A}), which is called {\it seed}, into another solution, possibly physically inequivalent to the initial one.
In practice the Harrison transformation (\ref{harrison}) acts on an initial seed written in terms of the complex Ernst potentials $(\Er,\mathbf{\Phi})$ and maps it into another couple of Ernst potentials $(\bar{\Er},\bar{\mathbf{\Phi}})$. This process is able to generate new solutions without integrating the field equations, which is a non-trivial fact because they are a set of coupled partial differential equations.\\
Actually Hauser and Ernst have proven the Geroch conjecture, which states that any axisymmetric and stationery solution of General Relativity can be generated from the symmetries of the Ernst equations  \cite{ernst-hauser}.\\
In particular, the Harrison transformation is able to add an electromagnetic field to a seed spacetime. For instance it can add monopolar electric charge to the Schwarzschild black hole to transform it into the Reissner-Nordstrom solution\footnote{See section \ref{repara} and appendix B of \cite{many-rotating} for more details and for the proof that Harrison includes the Kramer-Neugebauer transformation. The Harrison transformation can be used also, but in a different setting, to embed the seed spacetime in the electromagnetic Melvin universe \cite{ernst-magnetic}.}. \\
Also Ernst in \cite{ernst-remove} applied the Harrison transformation to the accelerating Reissner-Nordstrom black hole to obtain a charged C-metric in a Melvin fluxtube. The difference with respect to the present work is that Ernst did not act with the Harrison transformation on the charged C-metric cast in the Lewis-Weyl-Papapetrou metric (\ref{LWP-metric}), but on its conjugate version, obtaining a completely different action of the Harrison transformation. For more information about the relation between the Lewis-Weyl-Papapetrou metric and its conjugate (or double Wick rotation) in the context of Ernst generating technique, see \cite{swirling}.  \\

\section{Reissner-Nordstrom black hole in a charged Rindler background}
\label{sec:generation}

\subsection{Generation of the new solution through the Harrison transformation}

Our objective is to build a black hole solution in an electromagnetic Rindler background, a generalisation of the C-metric, where the accelerating horizon carries some extra features. In order to do so we start choosing, as initial seed, the accelerating Reissner-Nordstrom black hole which can be written in terms of the Lewis-Weyl-Papapetrou (LWP) metric in spherical-like coordinates ($t,r,x=\cos\theta,\varphi$)\footnote{The range of angular coordinates is $x\in[-1,1]$ and $\varphi \in [0,2\pi]$.} as
\beq \label{lwp-rx}
        ds^2 = -f(r,x) \left[ dt - \omega(r,x) d\varphi \right]^2 + \frac{1}{f(r,x)} \left[ e^{2\gamma(r,x)}  \left( \frac{{d r}^2}{\Delta_r(r)} + \frac{{d x}^2}{\Delta_x(x)} \right) + \rho^2(r,x) d\varphi^2 \right] \ ,
\eeq
where\footnote{The $\gamma$ function considered here does not correspond exactly with the one in (\ref{LWP-metric}). In these coordinates it's more concise if $\gamma$ to absorb also part of the coordinate transformation $(\r,z) \to (r,x)$.} 
\bea
        f(r,x)      &:=& \frac{\D_r}{ r^2 \Om^2 }  \ \ , \\
        \om(r,x)    &:=& 0 \ \ , \\
        \g(r,x)     &:=& \mezzo \log \left( \frac{\D_r}{\Om^4} \right)  \  , \\
        \rho(r,x)   &:=& \frac{\sqrt{\D_r} \sqrt{\D_x}}{\Om^2} \ \ , \label{rho} \\
        \Delta_r(r) &:=& (1-A^2r^2) (r^2-2mr+e^2+p^2)  \ \ ,   \\
        \Delta_x(x) &:=& (1-x^2) [1+2mAx+A^2x^2(e^2+p^2)] \ , \\
        \Omega(r,x) &:=& 1 + A r x \ \ ,  \label{pd-end}
\eea 
and with the dyonic electromagnetic potential of the form
\beq
\label{A-Acc-RN}
       A_\m = \left( - \frac{e}{r} , 0, 0, p x  \right)   \ \ .
\eeq
The physical parameters $m,A,e,p$ are related to the mass, acceleration, electric and magnetic charge respectively. 
The transformation that has been used to pass from the Weyl cylindrical coordinates ($ \r , z $) to the spherical ones ($r,x$) is determined by ($\ref{rho}$) above and the remaining coordinate
\beq
          z(r,x) = \frac{(Ar+x)[r - m(1-Arx) - Ax (e^2+p^2)]}{\Omega^2} +z_0\ , 
\eeq
Thanks to the definitions of the Ernst complex potentials (\ref{def-Phi-Er})-(\ref{h-e}) we can derive\footnote{We recall the form of the gradient in these spherical-like coordinates $(r,x)$: $\overrightarrow{\nabla} f \propto \sqrt{\D_r} \overrightarrow{e}_r \p_r f + \sqrt{\D_x} \overrightarrow{e}_x \p_x f $.}
\bea
      \tilde{A}_\varphi (r,x) &=& - \frac{p}{r} \  \  , \\      
      h(r,x) &=&  h_0\  \  , \\
      \mathbf{\Phi} (r,x) &=& - \frac{e + i p}{r} \label{Phi0} \  \  , \\
      \Er (r,x) &=&  \frac{\Delta_r}{r^2 \Omega^2} -\frac{e^2+p^2}{r^2}  \  \  , \label{Er0}     
\eea
The action of the Harrison transformation (\ref{harrison}) on the seed potentials (\ref{Phi0})-(\ref{Er0}) generates the new solution, that can be explicitly written in terms of the Ernst complex fields $(\bar{\Er},\bar{\mathbf{\Phi}})$, as follows
\bea
      \bar{\Er}(r,x) &=&  \frac{q^2\Omega^2-\D_r}{s^2\D_r-[s^2q^2+2s(e+ip)r+r^2]\Omega^2} \ , \\ 
      \bar{\mathbf{\Phi}}(r,x) &=&  \frac{[sq^2+r(e+ip)]\Omega^2-s\D_r}{s^2\D_r-[s^2q^2+2s(e+ip)r+r^2]\Omega^2}    \  \ .
\eea
Note that for simplicity we have chosen a real parameter to label the Harrison transformation (\ref{harrison}), that is $\a=s$ and that the integrating constant $h_0$ can be fixed to zero without losing physical generality, while $q:=\sqrt{e^2+p^2}$. An imaginary component in $\a$ would add a further magnetic field. \\ 
The above Ernst potentials already represent the sought solution, the accelerating Reissner-Nordstrom black hole endowed with a dyonic electromagnetic field and an additional electromagnetic charge encoded in $s$. By construction they fulfil all the Ernst equations (\ref{ee-ernst-ch})-(\ref{em-ernst}). However, in order to express the new solution in the metric and potential form, one has to appeal again to the definitions  (\ref{def-Phi-Er})-(\ref{h-e}). To keep the model as simple as possible, we chose to set the seed magnetic charge to zero, i.e. $p=0$, in any case the general solution for $p \neq 0$ can be found in appendix \ref{app:dyonic-acc-RN-I}. Vanishing $p$ also has the advantage of removing issues related with the Dirac string and magnetic monopoles. The line element after the transformation reads
\beq \label{lwp-rx-new}
        ds^2 = - \frac{f(r,x) \ dt^2}{|\L(r,x)|^2}  + \frac{|\L(r,x)|^2}{f(r,x)} \left[ e^{2\gamma(r,x)} \left( \frac{{d r}^2}{\Delta_r(r)} + \frac{{d x}^2}{\Delta_x(x)} \right) + \rho^2(r,x) d\varphi^2 \right] \ .
\eeq
So basically in the metric only $f$ change by  rescaling by a factor $|\L|^2=|1 -2s\mathbf{\Phi} - s^2 \Er|^2$; while the only non-null component of the electromagnetic vector potential is the electric one
\beq \label{At}
              A_t = \frac{\mathbf{\Phi} + s \Er}{\L} = \frac{s(r-2m)(1-A^2r^2)-e(1+Arx)^2-Ase^2[2x+Ar(1+x^2)]}{r+s[2e+A^2se^2r-s(r-2m)(1-A^2r^2)]+Ax(2+rAx)(se+r)^2}  \ .
\eeq
The metric remains diagonal in this case, but as can be seen in the appendix \ref{app:dyonic-acc-RN-I}, when the intrinsic magnetic charge of the seed black hole is non null, for $p \neq 0$, it couples with the electric charge of the charged Rindler background generating a stationary rotation, thus $\omega(r,x) \neq 0$ after the Harrison transformation. That's a manifestation of the Lorentz force: the generalised Reissner-Nordstrom black hole rotates, but with zero angular momentum. The rotation is therefore not encoded in the angular dipole, but in the subsequent angular multipole moments, which in general do not necessary contribute to the angular momentum\footnote{It may seems counter-intuitive because the most famous rotating solution is probably the Kerr metric, where the angular multipole moments are all determined as functions of the angular momentum, therefore switching off one angular multipole is sufficient to switch all of them. However this is not the case for more general solutions: multipole moments of different orders can be independent.}. This kind of behaviour  resemble what happens with magnetised (accelerating) black holes \cite{ernst-magnetic}, \cite{marcoa-pair}, where we can have a rotating (accelerating) Reissner-Nordstrom black hole which rotate because of the interaction with the electromagnetic background.\\  
The greatest peculiarity of this spacetime described in eqs. (\ref{lwp-rx-new})-(\ref{At}) is that it has two independent electric charges: $e$ and $s$. In the limit of vanishing $s$ we recover the seed, that is the standard accelerating Reissner-Nordstrom black hole, but for $s\neq 0$ (even if $e=0$) we have a novel spacetime, which still represents an accelerating and charged black hole! First of all, to prove that the novel metric is inequivalent with respect to the known accelerating Reissner-Nordstrom one, it is sufficient to check if it does not belong to the same type according to the Petrov classification.\\  

\subsection{Petrov Type I}

It well known that the standard accelerating Reissner-Nordstrom black hole, $s=0$, in our solution (\ref{lwp-rx-new})-(\ref{At}) is part of the Plebanski-Demianski family of spacetimes, therefore it has to belong to the type-D of the Petrov classification. \\
If the new accelerating and charged Reissner-Nordstrom black hole does not fall into the D-type, then the generated metric can not be just a diffeomorphism of the standard case.  So we compute the scalar invariants related to the Weyl tensor to check the algebraic class of the solution (\ref{lwp-rx-new})-(\ref{At}). In particular we focus on the scalar invariant 
\beq
\label{I327J2}
                     I^3 -27 J^2 \ ,
\eeq
where
\begin{equation}
I = \Psi_0\Psi_4 - 4 \Psi_1\Psi_3 + 3\Psi_2^2 \qquad , \qquad
J = \det
\begin{pmatrix}
\Psi_0 & \Psi_1 & \Psi_2 \\
\Psi_1 & \Psi_2 & \Psi_3 \\
\Psi_2 & \Psi_3 & \Psi_4
\end{pmatrix} \ .\\
\end{equation}
If the quantity in (\ref{I327J2}) is null the spacetime is algebraically special.
After evaluating the scalar invariant, we conclude that the spacetime generated in this section is, generically, not of type D, but algebraically general, that is of Petrov Type I. In fact apart for the seed case, for $s=0$, and some other know sub-cases (such as the non accelerating specialization $A=0$), the full solution is not of type D, neither for $e=0$ (and $p=0$). On the other hand the accelerating $s$-charged background, defined by the absence of the black hole $m=0$ (and $e=0=p$), remains of type D, that's because the scalar invariant in eq. (\ref{I327J2}) is null and also\footnote{If $I=J=0$, the vanishing of the scalar invariant in (\ref{I327J2}) could imply a metric of Type III or N, while if $I=J=0$ does not hold (but (\ref{I327J2}) holds) the spacetime could be II (if $K=N=0$) or else D, see fig 9.1 of \cite{stephani-big-book} for details.}
\bea
      I &\neq& J  \ = \ 0 \ , \\
      K &=& \Psi_1 \Psi_4^2 - 3 \Psi_4\Psi_3 \Psi_2 + 2 \Psi_3^3 \ = \ 0 \ , \hspace{3cm} N \  = \ 12 L^2-\Psi_4^2 I  \ = \ 0 \ ,
\eea
with 
\beq
            L = \Psi_2 \Psi_4 -\Psi_3^2  \ .
\eeq
The definitions of the Newman-Penrose scalars $\Psi_i$ necessary to compute the above scalar invariants can be found below
\bea
        \Psi_0 &:=& C_{\m\n\s\r} k^\m m^\n  k^\s m^\r \ , \nn  \\
        \Psi_1 &:=& C_{\m\n\s\r} k^\m l^\n  k^\s m^\r \ ,  \nn  \\
        \Psi_2 &:=& C_{\m\n\s\r} k^\m m^\n  \bar{m}^\s l^\r \ ,  \\
        \Psi_3 &:=& C_{\m\n\s\r} l^\m k^\n  l^\s \bar{m}^\r \ , \nn \\
        \Psi_4 &:=& C_{\m\n\s\r} l^\m \bar{m}^\n  l^\s \bar{m}^\r  \ . \nn
\eea
These five complex scalar functions characterise the Weyl tensor. They can be explicitly computed after defining a null Newman-Penrose tetrad. We have choosen the following tetrad
\bea \label{tetrad}
         \bf{k} &=&  \left( \frac{1}{\sqrt{-2g_{tt}}} \ \p_t + \frac{1}{\sqrt{2g_{xx}}} \ \p_x \right) \ , \\
         \bf{l}  &=&  \left( \frac{1}{\sqrt{-2g_{tt}}} \ \p_t - \frac{1}{\sqrt{2g_{xx}}} \ \p_x \right) \ , \\
         \bf{m}   &=&  \left( \frac{g_{t\varphi}}{\sqrt{2Dg_{tt}}} \ \p_t + \frac{i}{\sqrt{2g_{rr}}} \ \p_r + \sqrt{\frac{g_{tt}}{2D}} \ \p_\varphi \right) \ ,
\eea
where
$$   D = g_{tt} g_{\varphi\varphi} - g_{t\varphi}^2  \ \ .  $$
The non null scalar products between these vectors are just $ k_\m l^\m =-1$ and $\ m_\m \bar{m}^\m =1 $. \\

\subsection{Conical singularities}

It is well known that accelerating black hole metrics generically present conical singularities on the symmetry axis. We inspect possible conicity by taking the ratio between the a small circumference around the azimuthal semi-axes $z$ (along both north and south directions, i.e. $\theta=\pi$ and $\theta=0$), and its radius. If this ratio is equal to $2\pi$ the spacetime is free from angular deficit or excess. In the set of coordinates\footnote{For further details about the conical singularity in these coordinates see \cite{marcoa-thermo}.} we are using here this quantity can be computed by the following limits
\bea
\label{con-sing-N}
   \frac{\text{north circumference}}{\text{radius}} \ = \   \lim_{x \to 1} \int_0^{2\pi}\frac{1}{1-x^2} \ \sqrt{\frac{g_{\varphi\varphi}}{g_{xx}}} \ d\varphi &=&  2\pi  \ (1+A^2e^2+2Am) \ , \\
\label{con-sing-S}    \frac{\text{south circumference}}{\text{radius}} \ = \   \lim_{x \to -1} \int_0^{2\pi}\frac{1}{1-x^2} \ \sqrt{\frac{g_{\varphi\varphi}}{g_{xx}}} \ d\varphi &=& 2\pi \ (1+A^2e^2-2Am)   \ .
\eea 
Even though the parametrization could be improved and we have a certain freedom in the range of the azimuthal angle $\varphi$, we can already appreciate from eqs. (\ref{con-sing-N})-(\ref{con-sing-S}) how the only way both the above limits are equal to $2\pi$ is for the trivial non accelerating case $A=0$, or the equally trivial no black hole seed mass case $m=0$. Therefore, as the standard C-metric these {\it exotic} C-metrics described in (\ref{lwp-rx-new})-(\ref{At}) still seem plagued by non removable conical singularities. Anyway we will come back to this computation afterwards, when in possession of a better physical parametrization, in section \ref{e0} and \ref{new-par-sol+e}. As for the standard case, by properly rescaling the coordinate $\varphi$, we can surely remove the north pole nodal singularity or alternatively the one on the south. Eventually the introduction of external fields, as done in \cite{ernst-remove}, \cite{marcoa-thermo} or \cite{marcoa-removal}, can provide a source for the acceleration and so it can cure both conicity simultaneously to remain with a regular metric and a completely smooth manifold, outside the event horizon.\\
Of course the spacetime considered in this section is diagonal, therefore this subcase with $p=0$ is not plagued with NUTty singularity or Misner strings. However the more general case of the full solution in appendix \ref{app:dyonic-acc-RN-I} can be affected both by Misner and Dirac strings. \\

\subsection{Reparametrization}
\label{repara}

Usually, after the Harrison transformation, a reparametrization of the solution in order to better appreciate its physical characteristics is useful. As discussed above, in the case of no acceleration, the solution remains of type D, therefore, because of the black hole uniqueness theorems, can not be anything different from the only asymptotically flat charged black hole of the Einstein-Maxwell theory: Reissner-Nordstrom. To see it explicitly we can do, in the solution (\ref{lwp-rx-new})-(\ref{At}), the following change of radial coordinate, a time rescaling and some reparametrization of the integrating constant as follows
\bea 
\label{change-coord1}
       r &\to &  \frac{\bar{r}(1-s^2)+2s(s\bar{m}-\bar{e})}{(1-s^2)^2}  \ , \hspace{1.4cm} m \ \ \to \ \ \frac{\bar{m}(1+s^2)-2s\bar{e}}{(1-s^2)^2} \ , \\
\label{change-coord2}      t & \to & \bar{t} \  (1-s^2)   \ , \hspace{3.7cm} e \ \ \to \ \  \frac{\bar{e}(1+s^2)-2s\bar{m}}{(1-s^2)^2} \ .
\eea
The non-accelerating solution then becomes precisely the electric RN black hole, whose metric and electric potential are respectively
\beq \label{standard-RN-g}
      ds^2 = - \left(1-\frac{2\bar{m}}{\bar{r}}+\frac{\bar{e}^2}{\bar{r}^2} \right) d\bar{t}^2 + \frac{d\bar{r}^2}{\left(1-\frac{2\bar{m}}{\bar{r}}+\frac{\bar{e}^2}{\bar{r}^2} \right)} + \frac{\bar{r}^2 dx^2}{1-x^2} + \bar{r}^2 (1-x^2) d\varphi^2 \ ,
\eeq
\beq \label{standard-RN-A}
          A_\m = \left( s - \frac{\bar{e}}{\bar{r}}, 0, 0 , 0 \right) \ .
\eeq
Thus the Harrison transformation, in this non accelerating case, acts practically as an identity, mapping RN in itself. The electric field brought by $s$ is not independent of the seed one: the two charges are basically the same. Note, however, that in case the seed would be uncharged, that is if we would have applied the Harrison to the Schwarzschild black hole, i.e. $e=0$, we would still have obtained the RN solution, in that case the parameter introduced by the Harrison transformation would have been the only and total electric charge, hence non-trivial, see section \ref{schw-charged-back} for more details. In that case (for $\ e=0$) the Harrison would not have been an identity.
We will see as, in the presence of the acceleration, the charge parameter $s$ is, in general, independent on the seed one ($e$).\\
The choice of the new parameters $\bar{m}$ and  $\bar{e}$ is significant also when the acceleration is non null as we will see below computing the electric charge of the black hole and inspecting the position of the horizons.\\ 
It's easy to realise that the reparametrization (\ref{change-coord1})-(\ref{change-coord2}) consistently reduces to the old one for vanishing $s$: $\bar{m} \to m,$ $\bar{r} \to r$, ...  Moreover it does not change a possible regularisation of the conical singularity, as found in (\ref{con-sing-N})-(\ref{con-sing-S}), for real values of the constants and under the requirement of preserving the event horizon, which means $\bar{m}>\bar{e}$. For instance the constraint for having a symmetric conicity on the axis (necessary prerequisite to regularise the both conical excesses or defects thanks also to the freedom in the $\varphi$ angle range) gives
\beq
         s = \frac{ \bar{e} \pm \sqrt{\bar{e}^2-\bar{m}^2}}{\bar{m}^2} \ ,
\eeq
which is incompatible with the presence of an non-extremal event horizon. Indeed, in the presence of acceleration, the horizons are determined by the loci of the function $\D_r$. In the new parametrization it means the inner and outer horizons and the accelerating horizons are 
\beq \label{rpm}
          \bar{r}_\pm = \bar{m} \pm \sqrt{\bar{m}^2-\bar{e}^2} \ , \hspace{3cm} \bar{r}_{A\pm}= \frac{\pm 1 \pm s^4 \mp 2As\bar{e}\mp 2s^2(1+A\bar{m})}{A(1-s^2)}
\eeq
A proper reparametrization of $s$, as below, better clarifies also the position of the accelerating horizon. Anyway it depends on the case under consideration.\\

\subsection{e = 0 : Discharged seed}
\label{e0}

It is illustrative to write the complete metric in the new parametrization. In order to keep the procedure as simple as possible we, for the moment, we leave the initial charge of the black hole $e=p=0$. In this way the formal analogy with the Ehlers transformation and the accelerating NUT case is even more apparent\footnote{That's because the seed used to generate accelerating Taub-NUT or the accelerating Reissner-Nordstrom black hole cannot carry NUT charge.}, \cite{Podolsky-nut}, \cite{PD-NUTs}. \\
When the intrinsic charge of the seed is zero, the parametrization (\ref{change-coord1})-(\ref{change-coord2}) becomes
\bea \label{new-para-e0}
           r &\to &  \frac{(\bar{r}-\bar{r}_-)\bar{r}_+}{\bar{r}_+-\bar{r}_-}   \ , \hspace{1.4cm} m \ \ \to \ \frac{\bar{r}_+}{2}  \ , \hspace{1.4cm} A \ \ \to \ \ 
          \left(1-\frac{\bar{r}_-}{\bar{r}_+} \right)  \bar{A}\ , \nn \\ t & \to &   \left(1-\frac{\bar{r}_-}{\bar{r}_+} \right)  \bar{t}     \ , \hspace{1.5cm}              s \ \ \to \ \sqrt{\frac{\bar{r}_-}{\bar{r}_+}} \ .     
\eea
Note that now $s$ is not free but is fixed in terms of the electric charge $\bar{e}$. In particular when the electric charge vanishes, $s=0$ as expected, and we come back to the seed. Also we added a convenient rescaling of the acceleration parameter $A$.
Thanks to this coordinate and parameters modification the metric takes the following simple form\footnote{A Mathematica file containing this solution is available between  the sources of the arXiv files, for the reader convenience.}
\beq \label{gdd-new-para-e0}
      ds^2 = \frac{1}{\Om^2} \left[ - \frac{\D_r}{\mathcal{R}^2}  d\bar{t}^2 + \frac{\mathcal{R}^2}{\D_r} d\bar{r}^2 + \mathcal{R}^2 \left( \frac{dx^2}{\D_x} + \D_x d\varphi^2 \right) \right] \ \ ,
\eeq
where 
\bea
       \Om(\bar{r},x)    &=&      1-  \bar{A} (\bar{r} - \bar{r}_-)x  \ ,  \\
       \D_r(\bar{r})  &=&  (\bar{r}-\bar{r}_+) (\bar{r}-\bar{r}_-) [1 - \bar{A}(\bar{r}-\bar{r}_-)]    \ ,    \\
       \D_x(x)             &=&     ( 1 - x^2) [1 - \bar{A}(\bar{r}_+ - \bar{r}_-)x]   \ ,  \\
       \mathcal{R}(\bar{r},x)  &=&   \frac{(\bar{r}-\bar{r}_-)\bar{r}_+}{\bar{r}_+ - \bar{r}_-} - \frac{\bar{r}_- \D_r}{(\bar{r}-\bar{r}_-)(\bar{r}_+-\bar{r}_-)\Om^2}   \ .  
\eea
The electric field that supports the above metric stems from the potential
\beq\label{Ad-new-para-e0}
           A_\m = \left\{\frac{\sqrt{\bar{r}_-/\bar{r}_+} (\bar{r}_- - \bar{r}_+) (\bar{r}-\bar{r}_+)[1-\bar{A}^2(\bar{r}-\bar{r}_-)^2] }{\bar{r}(\bar{r}_--\bar{r}_+) - 2 (\bar{r}-\bar{r}_-)^2 \bar{r}_+ x \bar{A} + (\bar{r}-\bar{r}_-)^2 [\bar{r}_+ \bar{r}_-(1+x^2)-\bar{r}(\bar{r}_- + \bar{r}_+ x^2)] \bar{A}^2} , \ 0 , \ 0 , \ 0 \right\} \ .
\eeq
For vanishing acceleration parameter, $\bar{A}=0$, the solution becomes the Reissner-Nordstrom black hole 
\bea
         ds^2 &=& - \frac{(\bar{r} - \bar{r}_+)(\bar{r} - \bar{r}_-)}{\bar{r}^2} d\bar{t}^2 + \frac{d\bar{r}^2}{(\bar{r} - \bar{r}_+)(\bar{r} - \bar{r}_-)} + \frac{\bar{r}^2  dx^2}{1-x^2} + \bar{r}^2 (1-x^2) d\varphi^2 \ , \\
         A_\m &=& \left( cost - \frac{\sqrt{\bar{r}_+ \bar{r}_-}}{\bar{r}},  \ 0 , \ 0 , \ 0 \right) \ \ .
\eea
While for vanishing the electric charge parameter $\bar{e}=0$ (or equivalently $\bar{r}_-$ and $\bar{r}_+=2m$), we have the accelerating Schwarzschild black hole, i.e. the standard C-metric
\beq
     ds^2 =  \frac{\displaystyle \left( 1 - \frac{\bar{r}_+}{\bar{r}}  \right) (1-\bar{A}^2 \bar{r}^2) dt^2 + \frac{d\bar{r}^2}{\left( 1 - \frac{\bar{r}_+}{\bar{r}}  \right) (1-\bar{A}^	2 \bar{r}^2)} + \frac{\bar{r}^2 \ dx^2}{(1-x^2)(1+\bar{A}\bar{r}_+ x)} + \bar{r}^2 (1-x^2)(1+\bar{A}\bar{r}_+ x) d\varphi^2}{(1-\bar{A} \bar{r} x)^2} .  \nn
\eeq
Interestingly enough the full solution is not the charged C-metric but still represents an accelerating Reissner-Nordstrom black hole. The difference between the two can be easily verified computing the scalar invariant which characterises the Petrov type. While the charged C-metric is type D, this black hole is of type I. \\
From a physical perspective we expect this solution to be a limit of a charged black hole binary system, as the one described in \cite{many-rotating}\footnote{Where the gravitational external fields are considered null, by switching off their respective parameters.}. In fact from one hand the standard C-metric is a limit of the Bach-Weyl binary system, for a limit where one of the black hole pair grows to infinity, while maintaining their distance unvaried \cite{bubble}. On another hand the Harrison transformation is known to equally charge to both the sources of a black hole Bach-Weyl binary \cite{many-rotating}. When on top of this picture we act on the C-metric with the Harrison transformation, we get the above solution, which therefore models a charged Reissner-Nordstrom black hole in a charged accelerating background. Actually it can be demonstrated, see section \ref{limit}, that this picture represents the zooming in the proximity of the event horizon of a binary system, where one of the two black holes is much larger with respect to the other. According to this perspective the near horizon limit focuses on just a portion of the black hole, which is encoded mathematically in the metric into the Rindler horizon. Obviously since both the black holes are charged, also the accelerating background must carry some information about that new charge, introduced by the Lie-point transformation, which is responsible for switching the solution to the general I type.  \\
 Note that, in the particular case under consideration in this subsection, the charge parameter is just one, so the electric charges of the background and of the black hole  are dependent. However in the more general solutions in eqs. (\ref{lwp-rx-new})-(\ref{At}) and (\ref{funct-eneq0-inizio})-(\ref{funct-eneq0-fine}) we have two electric charges, one for the black hole and the other for the accelerating background. The standard charged C-metric therefore is a subcase of the solution (\ref{lwp-rx-new})-(\ref{At}) (the standard type-D dyonic C-metric version is a subcase of (\ref{lwp-rx-new-complete})-(\ref{barAt_new-complete}) instead). To verify this point is sufficient to  vanish the electric charge of the accelerating background, which trivially means setting $s=0$, so we retrieve the accelerating Reissner-Nordstrom seed, as in eqs. (\ref{lwp-rx})-(\ref{A-Acc-RN}). \\
The fact that the charged binary system has a special case, the Majumdar-Papapetrou solution \cite{majumdar}, where it is possible to find an equilibrium configuration between the two black holes without introducing other external fields, suggests us to give another look to the conical singularity of the solution in this new parameterisation. Indeed when the binary sources are at equilibrium there are no conical excess or defects in the spacetime, that's because the gravitational attraction is balanced by the electromagnetic repulsion. If our physical interpretation is correct we should encounter a similar equilibrium configuration also in the type I accelerating and charged black hole. In the new parameterisation (\ref{new-para-e0}) the conical singularities on the north and the south pole are respectively
\bea
\label{con-sing-N-e0}
   \frac{\text{north circumference}}{\text{radius}} \ = \   \lim_{x \to 1} \int_0^{2\pi}\frac{1}{1-x^2} \ \sqrt{\frac{g_{\varphi\varphi}}{g_{xx}}} \ d\varphi &=&  2\pi  \ [1 - \bar{A} (\bar{r}_+ - \bar{r}_-)] \ , \\
\label{con-sing-S-e0}    \frac{\text{south circumference}}{\text{radius}} \ = \   \lim_{x \to -1} \int_0^{2\pi}\frac{1}{1-x^2} \ \sqrt{\frac{g_{\varphi\varphi}}{g_{xx}}} \ d\varphi &=& 2\pi \ [1 + \bar{A} (\bar{r}_+ - \bar{r}_-)]  \ .
\eea 
It is clear that both conical singularities are not present when the accelerating black hole is extreme, that is when $\bar{r}_+=\bar{r}_-$. That is not a case, indeed the extremality condition is the one which characterises the Majumdar-Papapetrou binary system. This point further strengthens the interpretation of these novel accelerating metrics as a part of a binary system. 
Anyway one has to be cautious regarding how to take this limit and its interpretation because it represents an extremal point also of the new barred coordinates and parameters. \\
Similar considerations will be done in the more general case where the seed is electrically charged $ e \neq 0$. Mapping the conical deficits or excess (\ref{con-sing-N})-(\ref{con-sing-S}) thanks to the new parametrisation (\ref{change-coord1})-(\ref{change-coord2}), we still formally obtain the extremality condition $\bar{m}=\bar{e}$\footnote{Even though now $\bar{m}$ and $\bar{e}$ depends also on the seed initial charge $e$.}.\\
Curvature scalar invariants for this spacetime display a divergent behaviour near the axis of symmetry, close to the conformal boundary, as observed for these kinds of metrics built in \cite{mann-stelea-chng}, so in general their physical interpretation as black holes is considered dubious. Note, however, that the generalisation constructed, by a combination of the Harrison and the Ehlers transformations, in section \ref{sec:generation-NUT} are free from this issue.\\

\subsection{Black hole electric charge}

We want to compute the black hole electric charge $Q$ for the general accelerating case (\ref{lwp-rx-new})-(\ref{At}). According to the Stokes theorem the charge passing through the boundary $\p\S$ of a spacelike hyper-surface $\S$ which surrounds the collapsed star\footnote{In order to have a well defined integral we consider that the hyper-surface $\S$ does not exceed the accelerating horizon and the conformal boundary.} is given by
\beq
         Q = - \int_{\p\S} d\varphi dx \sqrt{\hat{g}_{xx} \hat{g}_{\varphi\varphi}} \ n_\m \s_\n F^{\m\n} = \bar{e} \ ,
\eeq
where $n$ and $\s$ are respectively two time-like and spacelike unitary vectors normal to the boundary of $\S$, whose chosen normalisation is $n_\m n^\m = -1$ and $\s_\m \s^\m = 1$. With $\hat{g}_{ij}$ we call the induced metric on the bi-dimensional surface, for time and radial coordinate constant.
Thus the considered hyper-surface contains  both the electric charge $e$ of the seed and the one introduced by the Harrison transformation $s$. These two electric charges combine and contribute to the full electric charge, $\bar{e}$, of the newly generated accelerating black hole. This computation confirms the validity of the parametrization, substantially borrowed from the non-accelerating case (\ref{change-coord2}), to describe the physical properties of the black hole. \\
We stress that this is the electric charge of the black hole only, not of the whole spacetime containing also the (charged) accelerating background.\\
This specialization, for $p=0$, of the solution we are considering here has not magnetic field, clearly it has not magnetic charges, but the dyonic extension reported in appendix \ref{app:dyonic-acc-RN-I} carries magnetic charges.\\

\subsection{e $\neq$ 0: Two independent electric charges} 
\label{new-par-sol+e}

When we leave non-zero electric charge in the seed spacetime, the generated solution (\ref{lwp-rx})-(\ref{A-Acc-RN}) is endowed with two independent electric charges, the one of the seed $e$ and the one brought by the Harrison transformation, encoded in $\bar{e}$ (or alternatively $s$). According to the the new coordinates (\ref{change-coord1}) - (\ref{change-coord2}) the spacetime can still be written as in eq (\ref{gdd-new-para-e0}) where the metric functions are\footnote{A Mathematica file containing this solution is available between  the sources of the arXiv files, for the reader convenience.}
\bea \label{funct-eneq0-inizio}
       \Om(\bar{r},x)    &=&      1 +  \bar{A} x (\bar{r} - \hat{r}_-)   \ ,  \\
       \D_r(\bar{r})  &=&  (\bar{r}-\bar{r}_+) (\bar{r}-\bar{r}_-) [1 - \bar{A}(\bar{r}-\hat{r}_-)]    \ ,    \\
       \D_x(x)             &=&     ( 1 - x^2) [1 + \bar{A}x(\bar{r}_- - \hat{r}_-)] [1 + \bar{A}x(\bar{r}_+ - \hat{r}_-)]   \ ,  \\
       \mathcal{R}(\bar{r},x)  &=&   \frac{(\bar{r}-\bar{r}_-)\bar{r}_+}{\bar{r}_+ - \bar{r}_-} - \frac{\bar{r}_- \D_r}{(\bar{r}-\bar{r}_-)(\bar{r}_+-\bar{r}_-)\Om^2}   \ , \label{funct-eneq0-fine}
\eea
the Harrison parameter $s$ is fixed as function $\bar{r}_{\pm}, \hat{r}_-$ as follows
\beq \label{s}
         s =  \frac{\hat{r}_-}{\mp\sqrt{(\bar{r}_-- \hat{r}_-)(\bar{r}_+-\hat{r}_-)}+\sqrt{\bar{r}_+ \bar{r}_-}} \ .
\eeq
While the electric potential that generalise the one in (\ref{Ad-new-para-e0}) is
\beq
       A_t(\bar{r},x) = \frac{s(1-s^2)\D_r + (\bar{r} - s \sqrt{\bar{r}_+ \bar{r}_-})[s(\bar{r}_--\bar{r}_+) - (1+s^2) \sqrt{\bar{r}_+ \bar{r}_-} ] \Om^2}{(\bar{r} - s \sqrt{\bar{r}_+ \bar{r}_-})^2 \Om^2 -s^2\D_r} \ .
\eeq
In case we want to express the solution in terms of the physical parameters $e, \bar{e}, \bar{m}$, we just need the expression for the position of the inner and outer horizon $\bar{r}_{\pm}$, which remains formally defined as in (\ref{rpm}), even though now the the seed charge $e$ modifies $\bar{e}$, and it convenient to define a new quantity\footnote{The similitude with the accelerating black hole with two NUT charges of \cite{PD-NUTs} is apparent also in the parameterisation. In that case the seed charge was $\ell$, the charge introduced by the Lie-point transformation was $n$ and the total charge was $ n-\ell $. These quantities correspond to $e, \ \bar{e} - e $ and $ \bar{e} $ respectively. In this sense $\hat{r}_-$ has the same parametrization of the double nutty one of \cite{PD-NUTs}.}
\beq \label{rhat-}
           \hat{r}_- := \bar{m} - \sqrt{\bar{m}^2  - \bar{e}^2 + e^2 } \ .
\eeq
On the other hand the accelerating horizon in the new parameterisation takes a clearer form with respect to (\ref{rpm}):
\beq
              \bar{r}_A = \frac{1}{\bar{A}} + \hat{r}_- \ \ .
\eeq
Note that this location of the accelerating horizon holds also for the uncharged seed case, $e=0$, of (\ref{gdd-new-para-e0})-(\ref{Ad-new-para-e0}), but there $\hat{r}_- \to \bar{r}_-$. From the general case, the specific requirement $|\bar{e}|=|e|$ means that the total charge of the back hole after the Harrison transformation coincides with the initial charge of the seed, hence the charging transformation has left unchanged the electric charge, so the Harrison transformation becomes the identity map. In fact, in that case, it is easy to infer from (\ref{rhat-}) and (\ref{s}) that $s=0$, thus we remain with the type-D seed, as indicated also by the fact that the accelerating horizon is not shifted by a factor $\hat{r}_-$, since $\hat{r}_-$ vanishes.\\
On the other end, for vanishing the acceleration parameter $\bar{A}$, the solution becomes just the standard Reissner-Nordstrom (with electric charge shifted $\bar{e}$), as in (\ref{standard-RN-g})-(\ref{standard-RN-A}).\\
When the acceleration is non zero, the fact that $\bar{e} \neq e$ determines the switch of the Petrov to the general type I. Hence if $\bar{e}=0$, we still remain outside the type D, even though the metric describes a Schwarzschild black hole, but embedded in a charged accelerating background, as described below.\\
An interesting feature which comes with the presence of a double electric charge comes from the possibility of equilibrium configurations due to the interplay between the black hole charge and the electromagnetic field of the background, which might balance gravity. Actually in this case the analysis of the conicities, as defined in eqs. (\ref{con-sing-N-e0})-(\ref{con-sing-S-e0}), on the north and on the south poles gives respectively
\bea \label{con-2eN}
           \D_{\varphi_N} = \left[ 1 + e^4\bar{A}^4 +4\bar{A}^2 (\bar{m}^2-\bar{e}^2) + 4 \bar{A} \sqrt{\bar{m}^2-\bar{e}^2 + e^2}  + 8 e^2\bar{A}^2 (3-2\sqrt{\bar{m}^2-\bar{e}^2 + e^2}) \right] \ , \\  
     \label{con-2eS}       \D_{\varphi_S} = \left[ 1 + e^4\bar{A}^4 +4\bar{A}^2 (\bar{m}^2-\bar{e}^2) - 4 \bar{A} \sqrt{\bar{m}^2-\bar{e}^2 + e^2}  + 8 e^2\bar{A}^2 (3-2\sqrt{\bar{m}^2-\bar{e}^2 + e^2}) \right]  \ .
\eea
When $\D_\varphi \neq 2\pi$, the manifold becomes singular for the presence of conical excesses or deficits. In general, usual C-metrics, whether charged or not, cannot avoid axial singularities in both the north and south hemisphere of the event horizon. Nevertheless, as can be seen from (\ref{con-2eN})-(\ref{con-2eS}), the double charged accelerating black hole has the notable property that both the conical defects can be removed from its geometry, as announced in the $e=0$ case, because of the interaction between the charged black hole and the electrical background. A necessary condition for eliminating both conical singularities simultaneously is that  $\D_{\varphi_N} = \D_{\varphi_S}$, which means
\beq \label{cond-no-cony}
                 \sqrt{\bar{m}^2-\bar{e}^2 + e^2} \ \ \bar{A} (1 + e^2 \bar{A}^2) = 0  \ .
\eeq
Apart from the trivial cases, when there is no black hole or no acceleration, which are naturally regular, conical singularities can be removed for $\bar{m}^2-\bar{e}^2 + e^2=0$. However, because the event horizon condition requires $|\bar{m}|\geq \bar{e}$ and because by construction $|\bar{e}|\geq|e|$, this constraint (\ref{cond-no-cony}) can be accomplished only at extremality for uncharged seed ($e=0$), as we have already seen in section \ref{e0}.\\
From the careful analysis of scalar invariants we noticed that the curvature can become unbounded between the outer and the accelerating horizons, when the Harrison charge is present. Nevertheless the generalisations of sections  \ref{sec:generation-NUT} and \ref{app:full-PD} cure this defect, because there the NUT parameter is also introduced.

\subsection{Extremal near horizon metric and CFT entropy}

It is worth studying the near horizon behaviour of the above metric in particular to clarify some basic characteristics of the new accelerating black hole of the above section. In fact in zooming near the horizon we focus more on the black hole properties with respect to the background. \\
As explained in \cite{compere-kerr-cft} to describe the metric close by the extremal event horizon $r_e$ we adapt the coordinates ($t,\bar{r}$) as follows
\beq \label{change-coordinate}
           \bar{r}(\tilde{r}) := r_e + \varepsilon r_0 \tilde{r}    \quad  ,  \hspace{2cm}      t(\tilde{t}) := \frac{r_0}{\ve} \tilde{t}     \quad  ,  \nn 
\eeq
where the constant $r_0$ is needed to compensate the overall scale of the near-horizon geometry. Also the electric potential has to be adjusted by a proper constant  $\Phi_e = -\chi^\m A_\m |_{\bar{r}=\bar{r}_+}$ before the near horizon limit, in this way
\beq
              A_t \to A_t + \Phi_e \ .
\eeq
The near horizon limit for the extremal metric is obtained for $\ve \to 0$. As expected by the results of \cite{lucietti-kunduri}, it can be written as a warped product of $AdS_2 \times S^2$
\beq \label{metric-near}
           ds^2 = \G(x) \left[ -\tilde{r}^2 dt^2 + \frac{d\tilde{r}^2}{\tilde{r}^2} + \a^2(x) \frac{dx^2}{1-x^2} + \g^2(x) \ d\varphi^2 \right] \quad ,
\eeq 
where  
\bea \label{fields1}
           \G(x) &=& \frac{\bar{e}^2}{(1-A^2e^2)(1+Aex)^2} \hspace{1.6cm} , \hspace{1.8cm} r_{0} \ = \ \frac{\bar{e}}{\sqrt{1-e^2A^2}}  \ \ ,    \\
    \label{fields2}  
    \g(x)  &=&  \sqrt{1-x^2} \sqrt{1-e^2A^2} (1+exA) \ \quad , \qquad \hspace{0.9cm} 
  \a(x) \ = \   \frac{\sqrt{1-A^2 e^2}}{1+Aex}   \ . 
\eea
The near horizon electric potential goes as 
\beq
       A = \e \tilde{r} dt \ ,
\eeq
with 
\beq
             \e = \frac{\bar{e}}{1-A^2 e^2}   \ .
\eeq
This is a typical near horizon behaviour for an extremal charged accelerating black hole, as described in \cite{acc-cft}. This behaviour in proximity of the event horizon allows us to use the tool provided by the Kerr/CFT correspondence \cite{strominger-kerr-cft}, \cite{strominger-duals}, \cite{compere-kerr-cft} to estimate the entropy of the black hole through the mapping with a dual conformal model living on the boundary of the near horizon metric.  \\
We briefly present here the relevant results for the metric under consideration, for more details about the relation between extremal accelerating black holes and CFT, see \cite{acc-cft}. \\
First of all, from the asymptotic symmetries of the near horizon metric, it is possible to extract one central charge of the conformal field theory living on the boundary of  (\ref{metric-near}):
\beq
            c_Q =  3 \e \int_{-1}^1 \frac{dx}{\sqrt{1-x^2}} \ \G(x) \a(x) \g(x) = \frac{6 \bar{e}^3}{(1-A^2e^2)^2} \quad .
\eeq
The microscopic entropy of the conformal field theory system living on the boundary of the extremal near horizon metric can be computed, from the Cardy formula, as follows 
\beq \label{Scft}
              \mathcal{S}_{CFT} = \frac{\pi^2}{3} c_Q T_L \ ,
\eeq 
where $T_L$ is the chemical potential, of the asymptotic conformal model, associated to the Frolov-Thorne vacuum\footnote{The standard Hawking temperature cannot be taken in consideration at this purpose because at extremality it vanishes. Also note that we have fixed for simplicity to $2\pi$ the period of the compact extra dimension $\psi$ necessary to export the Kerr/CFT analysis from a rotational five dimensional model to a electromagnetic four dimensional one.}, which for the metric 
 of section \ref{new-par-sol+e}, is
\beq
            T_L = \lim_{\bar{r}_+ \to r_e} \frac{T_H}{\Phi_e^{ext}-\Phi_e}  = \frac{1}{2 \pi \e}  \ .
\eeq
Finally using the eq. (\ref{Scft}) we obtain the microscopic entropy of the boundary conformal field theory
\beq
            \mathcal{S}_{CFT} = \frac{\bar{e^2 \pi}}{1-A^2e^2} \ .
\eeq
We confirm that this precisely corresponds with the Bekenstein-Hawking entropy of the extremal black hole of section \ref{new-par-sol+e}, which is a quarter of its event horizon area
\beq
           \mathcal{A}_{r_e}   = \int_0^{2\pi} d\varphi \int_{-1}^1 \sqrt{g_{xx}g_{\varphi\varphi}} \ dx = \frac{4 \bar{e^2 \pi}}{1-A^2e^2}  \ .
\eeq
\\

\subsection{Schwarzschild black hole in a charged accelerating Rindler background}
\label{schw-charged-back}

A remarkable sub-case we can extract from the double electrically charged solution of section \ref{new-par-sol+e}, consists in tuning the electric integration constants to vanish the net electric charge of the black hole: $\bar{e}=0$, or equivalently in the old parametrisation\footnote{In terms of the new parametrisation $s = \pm(\bar{m}-\sqrt{\bar{m}^2+e^2})/e$.}
\beq
\label{c-m-e}
           s = \frac{-m \pm \sqrt{m^2-e^2}}{e} \ ,
\eeq
without turning off the electric field in the full spacetime. This fine tune of the two electric charges implies that the inner and the outer horizon coincides. Thus we get a Schwarzschild-like black hole, immersed in an accelerating and charged Rindler-like background. The fact that $\bar{e}$ is null does not mean that the full electric field of the spacetime is null, but just that the charge parameter $s$ introduced by the Harrison transformation is not free, but instead it is a fine tuned function of the seed RN black hole physical parameters. This particular constraint (\ref{c-m-e}) implies that the Harrison transformation has discharged the black hole part of the seed (the accelerating charged black hole), while at the same time it has added an electric field to the Rindler background, which had no intrinsic electric charge before the transformation. In fact if we turn down the mass and electromagnetic charges of the seed we remain with a neutral Rindler spacetime.  On the other hand in case we turn off only the remaining electric charge $e$ we recover the standard C-metric, that is the accelerating Schwarzschild black hole.\\
It can be noted that, for small accelerations, the constraint (\ref{c-m-e}) also eliminates the characteristic monopole divergent term for $r=0$ of the Coulomb central potential. The small acceleration limit indeed means pushing the accelerating horizon to spatial infinity, thus we remain only with the standard Schwarzschild black hole.

\subsection{Charged C-metric in charged Rindler background as limit of a charged black hole binary system}
\label{limit}

\begin{figure}[h]
	\centering
	\includegraphics[scale=0.9]{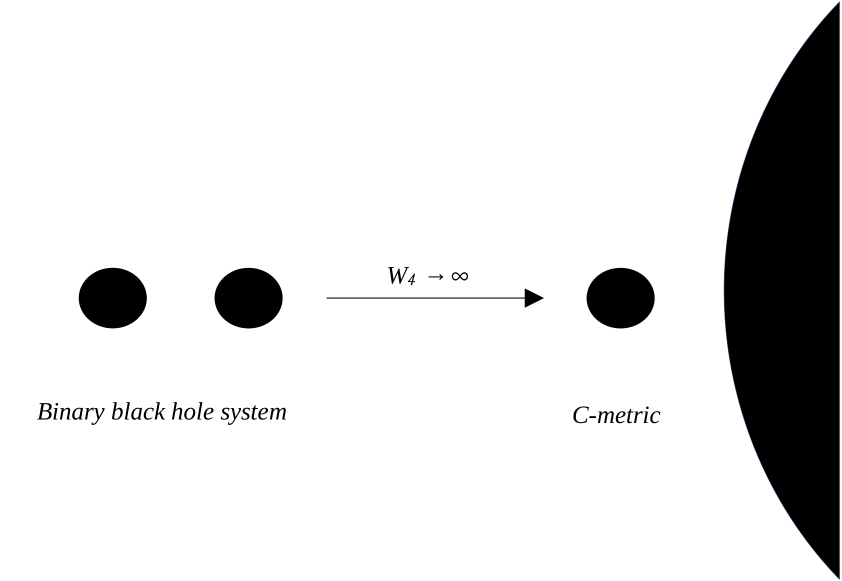}
	\caption{\small The limit for $w_4 \to \infty$ brings a binary black hole system into an accelerating black hole metric. For bigger values of $w_4$ the right element of the binary grows and its event horizon transforms into an accelerating horizon. If the right black hole of the binary carries no charges, then the metric after the limit becomes of special algebraic type D and the accelerating horizon is a standard Rindler one. If the right black hole of the binary carry some charges, such as the electric charge, the resulting accelerating metric remains of general type I and the Rindler background is endowed with that charge. Conical singularities here are not displayed, eventually they can be removed with external electromagnetic or gravitational fields in the spirit of \cite{ernst-remove} and \cite{many-rotating}.}
	\label{fig:limit}
\end{figure}

We would like here to provide some more elements on the interpretation, as announced in section \ref{e0}, of the new parameter, $s$ or $\bar{e}$, introduced by the Harrison transformation as an electric charge which affects also the Rindler background. Specifically we would like to obtain the type I charged C-metric as a limit of a charged binary system. \\
Consider a black hole binary system, such as the Bach-Weyl solution, which describes the simplest metric containing a couple of black holes in general relativity. In Weyl coordinates, which are much more economical for this setting, that metric takes the form
\beq
ds^2 = -\frac{\m_1 \m_3}{\m_2 \m_4} d \hat{t}^2  + \frac{16 C_f \m_1^3 \m_2^5 \m_3^3 \m_4^5 \  (d\r^2 + dz^2)}{\m_{12}\m_{14} \m_{23} \m_{34} W_{13}^2 W_{24}^2 W_{11} W_{22} W_{33} W_{44}}  + \r^2 \frac{\m_2 \m_4}{\m_1 \m_3}  d\hat{\varphi}^2  \ ,
\eeq  
where the fundamental blocks of the metric, as in Weyl or in the inverse scattering methods \cite{belinski-book} are the solitons, defined as
\beq
\m_i(\r,z) := w_i - z + \sqrt{\r^2 + (w_i-z)^2} \ ,
\eeq
while
\beq
\m_{ij}:= (\m_i -\m_j)^2  \ \ , \hspace{2.5cm} W_{ij}:=\r^2+\m_i\m_j \ .
\eeq
The physical quantities of the spacetime are encoded in the four $w_i$ parameters, which for instance to describe two Schwarschild black holes of mass $m_1$ and $m_2$ centred on the axis of symmetry at $z_1$ and $z_2$ are chosen as $w_1=z_1-m_1, w_2=z_1+m_1, w_3=z_2-m_2, w_4=z_2+m_2$. $C_f$ is just a gauge constant that can be adjusted at will, often to remove a conical singularity, whether it is present. \\ 
We can build the charged version of the Bach-Weyl metric thanks to the Harrison transformation (\ref{harrison}), where the electric parameter is $\hat{\a}$. The resulting solution is given by
\bea
ds^2 = - \frac{\m_1 \m_2 \m_3 \m_4}{(\m_2 \m_4 - \hat{\a}^2 \m_1 \m_3)^2} d\hat{t}^2 + \frac{16 C_f \m_1^3 \m_2^3 \m_3^3 \m_4^3 (\m_2 \m_4 - \hat{\a}^2 \m_1 \m_3)^2  \  (d\r^2 + dz^2)}{\m_{12}\m_{14} \m_{23} \m_{34} W_{13}^2 W_{24}^2 W_{11} W_{22} W_{33} W_{44}}   + \r^2  \frac{(\m_2 \m_4 - \hat{\a}^2 \m_1 \m_3)^2}{\m_1 \m_2 \m_3 \m_4}  d\hat{\varphi}^2  , \nn
\eea
\beq
A_\m = \left( A_{t_0} + \frac{\hat{\a} \m_1 \m_3}{\m_2 \m_4 - \hat{\a}^2 \m_1 \m_3} , 0 , 0 , 0 \right) \ .
\eeq
This kind of solution\footnote{Actually in \cite{many-rotating} a slightly different charging Lie-Point symmetry of the Ernst equations has been used, moreover it is embedded in an external multi-polar background (which here we are neglecting). However in \cite{many-rotating} it is shown as the Kramer-Neugebauer and the Harrison transformations are physically equivalent, because they differ only on gauge transformations.} contains the Majumdar-Papapetrou black hole pair at equilibrium, as a special extreme case, \cite{many-rotating}. \\
To show the relation with the type-I accelerating black hole we perform the following change of coordinates
\bea
\r & \to &  \frac{\sqrt{r(r-2m)(1-A^2r^2)(1+2Amx)(1-x^2)}}{(1+Arx)^2} \ , \hspace{1.6cm} \hat{t} \ \to \ \sqrt{\frac{A}{2w_4}} \ t  \ , \\  
z  & \to &  z_1 + \frac{(Ar+x)[r-m(1-Arx)]}{(1+Arx)^2} \ ,    \hspace{3.4cm} \hat{\varphi} \ \to \  \sqrt{\frac{2w_4}{A}} \ \varphi \ .
\eea
Then we center one black hole, let's say the left one, with mass parameter\footnote{Remember that the mass parameter before the Harrison transformation does not necessarily coincide with the mass of the black hole after the transformation.} $m$ in $z_1$, so $w_1 = z_1 -m$ and $w_2 = z_1 +m $. We fix just one physical parameter entering in the remaining couple of solitons, which will be related to the position of the accelerating horizon as $z_3=z_1+1/(2A)$, while $z_4$ remains free. Without losing generality we can assume that the origin of the $z$-axis is at $z_1$, thus we can set $z_1=0$. We fix also the arbitrary coefficient in the $g_{\r\r}$ element of the metric, such that  $C_f=2w_4m^2/A^3$ and we fix $\hat{\a}=\sqrt{A} s$. \\
Now if we take the limit for $w_4 \to \infty $, which in the Weyl representation means make the right black hole of the binary grow indefinitely bigger, see picture 2 of \cite{PD-NUTs}, we get
\beq \label{g-e0-inizio}
ds^2 = -f(r,x) dt^2 + \frac{1}{f(r,x)} \left[ e^{2\gamma(r,x)}  \left( \frac{{d r}^2}{\Delta_r(r)} + \frac{{d x}^2}{\Delta_x(x)} \right) + \rho^2(r,x) d\varphi^2 \right] \ ,
\eeq
where
\bea
f(r,x)      &:=& \frac{r^2\D_r\Om^2}{(s^2\D_r - r^2 \Om^2)^2 }  \ \    \hspace{2.2cm}    \Omega(r,x) \ := \ 1 + A r x \ , \nn \\
\g(r,x)     &:=& \mezzo \log \left( \frac{\D_r}{\Om^4} \right)  \ \ ,   \hspace{2.8cm}   \Delta_r(r) \ :=\  (1-A^2r^2) (r^2-2mr) \ ,\nn \\
\rho(r,x)   &:=& \frac{\sqrt{\D_r} \sqrt{\D_x}}{\Om^2} \ \ ,            \hspace{3.1cm}  \Delta_x(x) \ := \ (1-x^2) (1+2mAx)  \ ,  \nn
\eea
\beq
A_\m = \left(  \frac{s\D_r}{r^2\Om^2-s^2\D_r}  0, 0, 0 \right)  \ .  \label{A-e0-inizio}
\eeq
This coincides precisely with the solution of section (\ref{lwp-rx-new})-(\ref{At}) for $e=0$. Therefore we conclude that the Petrov type I charged and accelerating metric is the limit of a binary system where both black holes are charged. In the limiting process the black hole horizon of the right black hole grows so big, compared with the other, that it can be considered just a Rindler horizon\footnote{This is generalisation of the well known fact that the metric near a Schwarzschild black hole is the Rindler one.}, see picture \ref{fig:limit}. Since both the elements of the binary were charged the electric charge remains also as a feature of the accelerating background.

Note that when the electric charge $s$ is null, then this limit still works. In that case the limit for $ w_4 \to \infty $ of the Bach-Weyl binary black hole gives the standard type-D C-metric. \\
On the contrary applying this limit to more general binary black hole solution, such as the ones with two independent electric charges such as \cite{alekseev-belinski-2RN} or \cite{manko-2007}, one can retrieve also more general accelerating type I black holes with $e \neq 0 $, such as the full solution (\ref{lwp-rx-new})-(\ref{At}). In case we would like to extend this limiting procedure for the dyonic generalisation, when both   $e \neq 0 $ and $p\neq0$, as in \ref{app:dyonic-acc-RN-I}, or for NUTty backgrounds such the ones described in \ref{sec:generation-NUT}, since they are endowed with NUT parameter, it would be necessary even more general charged and rotating binary metrics, where also off diagonal terms are switched on. \\

\section{Reissner-Nordstrom-NUT in a charged Rindler-NUT background}
\label{sec:generation-NUT}

Recently a large class of type I black holes has been proposed in \cite{PD-NUTs}. This family consists of a NUTty extension of the Plebanski-Demianski black hole. A particular subcase of this very general spacetime is the accelerating Reissner-Nordstrom-NUT black hole, first constructed in \cite{tesi-giova}, which can be obtained from the general class of \cite{PD-NUTs}, by vanishing the angular momentum. We may think that if we act with the Harrison transformation (\ref{harrison}) to an accelerating NUTty black hole we might obtain the same solution described in \cite{PD-NUTs}. That's because both spacetimes represent charged and NUTty accelerating black holes and because they belong to the same Petrov class I. Nevertheless these two spacetimes are different. The main difference consists in the background: the case studied in \cite{PD-NUTs} represents an accelerating Reissner-Nordstrom-NUT black hole in a NUTty Rindler background while the solution we will construct here below, thanks to the extra Harrison transformation, models an accelerating RN-NUT black hole but in a NUTty and charged Rindler background. Let's further clarify this point: when we remove the black hole (with his charges) from the first case we remain with a Rindler-NUT spacetime, while in case we remove the black hole in the second spacetime, we are left with a electromagnetic Rindler-NUT background. In the former case we have a stationary rotating background without an electromagnetic field, in the latter a background with a non-null electromagnetic field. So these two solutions cannot be physically equivalent. \\

\subsection{Harrison and Ehlers transformations commute}

We recall that the accelerating RN-NUT black hole built in \cite{tesi-giova} and its rotating generalisations \cite{PD-NUTs} were generated by the Ehlers symmetry of the Ernst equations (\ref{ee-ernst-ch})-(\ref{em-ernst}), that is
\beq
         (III):  \qquad  \Er \longrightarrow \bar{\Er} = \frac{\Er}{1+ic\Er} \quad , \hspace{2cm}  \mathbf{\Phi} \longrightarrow  \bar{\mathbf{\Phi}} = \frac{\mathbf{\Phi}}{1+ic\Er} \ .
\eeq
It is a non-trivial fact that the Harrison (\ref{harrison}) and Ehlers $(III)$ transformations commute. That's because these Lie-point symmetries of the Ernst equations belong to the non commutative group SU(2,1). Nevertheless it is easy to check that
\beq \label{commute}
         (V) \circ (III) \ = \ (III) \circ (V) = \ \left\{\begin{matrix}
 \Er &\longrightarrow &  \displaystyle \bar{\Er} = \frac{\Er}{1 + i c \Er - \a \a^* \Er -2\a^*  \mathbf{\Phi}} \\
 & & \\
 \mathbf{\Phi} &\longrightarrow & \bar{\mathbf{\Phi}} = \displaystyle  \frac{ \mathbf{\Phi} + \a \Er }{1 + i c \Er - \a \a^* \Er -2\a^*  \mathbf{\Phi}}
\end{matrix}\right. \quad  .
\eeq
This observation opens to the possibility of generating a common solution which extends both the one built in this section and the one of \cite{PD-NUTs}. In fact starting with an accelerating solution as a seed, for instance the PD class (without the cosmological constant\footnote{It is well known that the symmetries of the Ernst equations in the presence of the cosmological constant are not effective to generate new solutions \cite{melvin-lambda}.}), it is possible to build a large class of accelerating, charged, with NUT parameter and endowed with angular momentum in an electromagnetic and  NUTty Rindler background. Basically because of the commutative property (\ref{commute}), we have just to apply $(V)$ to the general metric of \cite{PD-NUTs} to add electromagnetic charge to the Rindler background.\\
To keep the construction simple we start below with a diagonal seed, without gravitomagnetic mass nor angular momentum.

\subsection{Generation of a Reissner-Nordstrom-NUT black hole in a charged Rindler-NUT background} The above composition of Harrison and Ehlers transformations will be used to generate the accelerating Reissner-Nordstrom-NUT black hole in an electromagnetic Rindler-NUT background, that is the NUTty generalisation of the metric build in section \ref{sec:generation}. We start, as a seed, with the same accelerating RN black hole in (\ref{lwp-rx})-(\ref{A-Acc-RN}) and we apply the map (\ref{commute}). The composite transformation $(V) \circ (III)$ in principle is labelled by three real parameters ($c$ and the two components of the complex parameter $\a$), but for the sake of simplicity we consider $\a=s$ real. In the case of dyonic seeds, an extra imaginary part for the Lie parameter $\a$ might at most add a further contribution to the magnetic part of the Maxwell field, which however can be reabsorbed by an electromagnetic rotation transformation (\ref{I}). Thus an imaginary component for$\a$ does not affect the geometry of the spacetime. The proof that the real choice of $\a$ does not affect the generality of the resulting metric for dyonic solution can be found in appendix \ref{app:full-PD}.   \\
The procedure to generate the new metric is very similar to the one executed in section \ref{sec:generation}, the seed Ernst potentials are the same, so we do not repeat all the steps here but just state the final results. The new Ernst potentials, generated by the transformation $(V) \circ (III)$ are\footnote{A Mathematica file containing this solution is available between  the sources of the arXiv files, for the reader convenience.}
\beq \label{lwp-rx-new-charged-nut}
        ds^2 = - \frac{f(r,x)}{|\L(r,x)|^2} \big[dt - \bar{\omega}(r,x) d\varphi\big]^2  + \frac{|\L(r,x)|^2}{f(r,x)} \left[ e^{2\gamma(r,x)} \left( \frac{{d r}^2}{\Delta_r(r)} + \frac{{d x}^2}{\Delta_x(x)} \right) + \rho^2(r,x) d\varphi^2 \right]  ,
\eeq
where
\bea
       |\L(r,x)|^2 &=& \frac{[s^2\D_r-(r^2+2ers+q^2s^2)\Om^2]^2+4p^2r^2s^2\Om^4+4cprs\Om^2(\D_r-q^2\Om^2)+c^2(\D_r-q^2\Om^2)^2}{r^4\Om^4} \ , \nn \\
       \bar{\omega}(r,x) &=& \frac{1}{A\Om^2} \Big\{ 4Aps \Big[x(s^2-1)+2A[r(s^2-x^2)-ms^2(1-x^2)] +A^2x[r^2(s^2-x^2) - q^2 s^2 (1-x^2)] \Big] + \nn \\
                      &+&  2c \Big[ 1+2Ax(m+r+es) + A^2 \Big(r^2+4es(r-m) + 4mx^2(r+es)- q^2(1-x^2) \Big) \Big] + \nn \\
                      &+& 2A^3x\Big( r^2 (m+es) -q^2(r+es) (1-x^2) \Big) \Big\} + \omega_0 \ . \ \label{lwp-rx-new-charged-nut-fine}
\eea
The electromagnetic field supporting the metric become
\bea
    & \hspace{-2.5cm} \displaystyle \bar{A}_t =  \frac{-er^3\Om^4+r^2s\Om^2(\D_r-3q^2\Om^2)+(3ers^2-cpr)\Om^2(\D_r-q^2\Om^2)-s^3(\D_r-q^2\Om^2)^2}{|\L(r,x)|^2 r^4 \Om^4} \ , \nn \\
  & \hspace{-0.35cm} \displaystyle  \bar{A}_\varphi  =  A_{\varphi_0} + \frac{px(1-3s^2) -cex + A(ce+3ps^2)(2m+Aq^2x)(1-x^2) - Ar(2+Arx)(ce+p(3s^2-x^2)) }{\Om^2} -\bar{\om} \bar{A}_t \nn
\eea
Note that the presence of the acceleration in the seed is fundamental to obtain a new spacetime. Otherwise we would remain within the Carter-Plebanski class and the Harrison (or the Ehlers) transformation on an already charged (or NUTty) seed would act just as an identity, after a proper reparametrization of the solution. In particular in the case under consideration, in absence of acceleration, we would just find RN-NUT (as can be easily realised by taking the limit $A\to0$ of the above solution). \\  
So the main novelty of this solution comes from having simultaneously added both the NUT and the electromagnetic charges also to the accelerating background, and that is the reason why these metrics go beyond the Plebanski-Demianski family, but becomes of Petrov Type I\footnote{The evaluation of the Petrov class can be found exactly as in section \ref{sec:generation}. However it is not surprising having a general algebraic type, since even the $c=0$ subcase, as analysed in the previous sections, or for $s=0$ as in \cite{PD-NUTs}, was already of type I.}. \\
Further generalisations may include also in angular momentum or NUT charge in the seed. In appendix \ref{app:full-PD} the most general Plebanski-Demianski seed is embedded in this NUTty and charged Rindler background. A detailed map of the main solutions discussed in this article can be found in figure \ref{fig:grafico}.\\

\subsection{Electric accelerating background} 
\label{sec:back}

When we remove completely the RN-NUT black hole, which means setting to zero its characteristic parameters ($m,e,p$), we remain with a stationary rotating Rindler metric with an electromagnetic field, not with the usual Rindler metric, as in the type D accelerating black hole solutions. In any case the electromagnetic field can be removed thanks to its parameter $s$. In an equivalent way we may have just started from a no black hole spacetime as seed, but only with the Rindler metric and after the transformation (\ref{commute}) we

\newpage

\tikzstyle{rect} = [draw,rectangle, fill=white!20, text width =3cm, text centered, minimum height = 1.5cm,scale=0.9]
\begin{figure}[H]
\vspace{-1cm}

\begin{center}

\begin{tikzpicture}

\hspace{-0.1cm}    \node[rect,scale=1.2,label=above :{type I \quad \qquad \qquad (\ref{Ebarfull})-(\ref{Phisbarfull})},text width=4.5cm,node distance=1.5cm,line width=1.3pt](PD double nut){Plebanski-Demianski-NUTs in charged accelerating background};
    \node[rect,scale=1.2,anchor=north,label=above :{type I \quad \qquad \qquad (\ref{lwp-rx-new-charged-nut})-(\ref{lwp-rx-new-charged-nut-fine})},
    text width=4.5cm,below of=PD double nut,node distance=3.3cm,line width=1.3pt](acc kerr newman nut){Dyonic Reissner-Nordström-NUT in charged accelerating background};
    \node[rect,scale=1.1,anchor=north,label=above :{type I \qquad \qquad \cite{PD-NUTs}},text width=4.2cm, left of= acc kerr newman nut, node distance=6cm](PD-NUTs){Plebanski-Demianski with double NUT charges};
    \node[rect,scale=1.1,anchor=north,label=above:{type D \qquad \qquad \cite{Plebanski-Demianski}},text width=3.5cm, right of= acc kerr newman nut, node distance=6cm](PD){Standard Plebanski-Demianski};
    \node[rect,scale=1.2,anchor=north,label=above :{type I \qquad \qquad \qquad \qquad  \cite{tesi-giova}},text width=4.8cm, below left of=acc kerr newman nut,node distance=4.7cm](acc RN NUT){Dyonic Reissner-Nordsrom-NUT in Rindler};
    \node[rect,scale=1.2,anchor=north,label=above :{type I \qquad \qquad \qquad (\ref{lwp-rx-new-complete})-(\ref{barAt_new-complete})},text width=4.5cm, right of= acc RN NUT,node distance=6.6cm,line width=1.3pt](kerr newman nut){Dyonic Reissner-Nordstrom in charged accelerating background};
    \node[rect,scale=1.2,anchor=north,below of=acc RN NUT,node distance=3.2cm,label=above left:{\; type D},text width=4.8cm](RN NUT){Accelerating Reissner-Nordström (charged C-metric)};
    \node[rect,scale=1.2,anchor=north,below of=kerr newman nut,node distance=3.2cm,line width=1.3pt, label=above :{\; type I \quad \qquad \qquad (\ref{funct-eneq0-inizio})-(\ref{funct-eneq0-fine})},text width=4.4cm](RN charged RIndler){Reissner-Nordström in charged accelerating background};
\node[rect,scale=1.2,anchor=north,below left of=RN NUT,node distance=4.3cm, label=above :{\;  type D \qquad \qquad },text width=3.2cm](RN){Reissner-Nordström};
\node[rect,scale=1.2,anchor=north,below left  of=RN charged RIndler,node distance=4.4cm, label=above :{type D },text width=1.8cm](c){C-metric};
\node[rect,scale=1.2,anchor=north,below right of =RN charged RIndler  , node distance=4.4cm, line width=1.3pt,label=above :{\qquad type I \qquad  (\ref{c-m-e})},text width=2.7cm](sch-charg-rindler){Schwarzschild in charged accelerating background};
\node[rect,scale=1.2,anchor=north,below of = c  , node distance=3cm, label=above :{ type D \qquad \qquad \  },text width=1.7cm](Rindler){Rindler};
\node[rect,scale=1.2,anchor=north,below of = sch-charg-rindler  , node distance=3cm, label=above :{ type I \qquad (\ref{background-I-inizio})  },text width=1.9cm,line width=1.3pt](back){Charged accelerating background};

\draw[->] (PD double nut) -- node {no angular \  momentum}(acc kerr newman nut);
\draw[->] (acc kerr newman nut) -- node [right] {$s$ = 0} (acc RN NUT);
\draw[->] (PD double nut) -- node [right,near start] {\; \; \;$c$ = 0}(PD);
\draw[->] (PD double nut) -- node [right] {\; \; \ $s$ = 0}(PD);
\draw[->] (PD double nut) -- node [left] {\; $s$ = 0}(PD-NUTs);
\draw[->] (acc kerr newman nut) -- node [right] {\; \; $c$ = $0$}(kerr newman nut);
\draw[->] (PD-NUTs) -- node [right] {\; $a=0$}(acc RN NUT);
\draw[->] (kerr newman nut) -- node [ right] {$p$ = 0} (RN charged RIndler);
\draw[->] (acc RN NUT) -- node [right] {\; \; $c$ = 0} (RN NUT);
\draw[->] (RN charged RIndler) -- node [above] {$s$ = 0} (RN NUT);
\draw[->] (RN NUT) -- node [right] {$A=0$}(RN);
\draw[->] (RN charged RIndler) -- node [right,near start] {\; \; $\bar{A}=0$}(RN);
\draw[->] (RN charged RIndler) -- node [right] {\; \; $\bar{e}=e=0$}(c);
\draw[->] (RN charged RIndler) -- node [right, near start] {\ $\bar{e}=0$}(sch-charg-rindler);
\draw[->] (RN NUT) -- node [left,near start] {$e=0$ \qquad }(c);
\draw[->] (c) -- node [right] {$m=0$ \qquad }(Rindler);
\draw[->] (sch-charg-rindler) -- node [below] {$e=0$ \qquad }(c);
\draw[->] (sch-charg-rindler) -- node [right] {$m=0$ \qquad }(back);
\end{tikzpicture}
\vspace{0.1cm}
\caption{Map of the family of accelerating solutions in the Einstein-Maxwell theory containing a black hole. The new spacetimes presented in this article, not yet known in the literature, are emphasized in bold line rectangles. The new family of solutions are of Petrov Type I and carry up to two independent NUT parameters and two independent electric (possibly also two magnetic) charges, when the acceleration is non-zero. However, note that accelerating black holes in external electromagnetic fields such as \cite{ernst-remove} and \cite{marcoa-pair} or in external rotational field \cite{marcoa-removal}, are not included in this family, they belong to a different branch. \\
}
\label{fig:grafico}

\end{center}
\end{figure}

would have obtained a Rindler background endowed with an electromagnetic field, whose metric reads
\bea \label{g-backgr-D}
      ds^2  &=& - \hat{f}(r,x) \left[dt - \left(\frac{2Acr^2(1-x^2)}{(1+Arx)^2} + \om_0 \right) d\varphi \right]^2  \\
            &+&  \frac{1}{\hat{f}(r,x)} \left\{ \frac{1}{(1+Arx)^4} \left[ dr^2+ (r^2-A^2r^4) \left( \frac{dx^2}{1-x^2} + (1-x^2) d\varphi^2 \right) \right] \right\} \nn
\eea
with
\beq
       \hat{f}(r,x) := \frac{(1-A^2r^2)(1+Arx)^2}{c^2(1-A^2r^2)^2 + [(1+Arx)^2 - s^2 (1-A^2r^2)]^2}   \ .
\eeq
The electromagnetic field stems from the potential
\beq \label{A-backgr-D}
             A_\m  = \hat{f} \left\{ s - s^3 \frac{1-A^2r^2}{(1+Arx)^2} ,\ 0  ,\ 0 ,\ \frac{2cs[1+Ar(Ar+2x)][(1-A^2r^2)s^2-(1+Arx)^2]}{A(1+Arx)^4}  \right\} \ .
\eeq
where we just shift the constant $\om_0 \to -2c/A+\om_0$ to have a well defined zero acceleration limit, the Minkowski spacetime. So, also in this background case, the relevance of the accelerating parameter and the accelerating horizon in the novelty of the solution is manifest. \\
From (\ref{A-backgr-D}) we clearly see that when $s \to 0$ the electromagnetic field goes to zero, thus we remain with just the standard Rindler metric. On the other hand when the acceleration is zero also the electromagnetic field vanishes. \\
Combining the fact that the full solution can be derived from blowing up one constituent of a charged binary system, as seen  in section \ref{limit}, and that the background, in this frame of reference, consists in vanishing the seed black hole quantities ($m, e, p$), it is natural to hypothesize that the background may be derived from a limit of the Reissner-Nordstrom black hole itself. This is in agreement with the uncharged case where the Rindler horizon can be obtained as a limit of Schwarzschild black hole: indefinitely enlarging the event horizon. To verify this hypothesis we consider the RN black hole written in cylindrical coordinates ($\rho,z$)
\beq
          ds^2 = - \frac{(R_+ + R_-)^2-4(m^2-q^2)}{(2m+R_+R_-)^2} dt^2 + \frac{(2m+R_+ + R_-)^2}{4 R_+ R_-} (d\r^2 + dz^2) + \frac{\r^2(2m+R_+R_-)^2 \ d\varphi^2}{(R_+ + R_-)^2-4(m^2-q^2)}  \ , \nn
\eeq
\beq
         A_\m = \left[ \frac{q^2}{(2m+R_+R_-)^2}, 0, 0, 0 \right] \ ,
\eeq
where
\beq
             R_\pm = \sqrt{\r^2 + \left[\pm (z - z_1) + \sqrt{m^2-q^2}\right]^2} \ \ .
\eeq
The relation with the usual solution in spherical coordinates, as in (\ref{standard-RN-g})-(\ref{standard-RN-A}), is given by the following map
\bea
         \rho  &=& \sqrt{r^2-2mr-q^2} \ \sqrt{1-x^2} \ ,\\  \nn
         z      &=&  z_1 + (r-m) \ x  \ .\nn
\eea
The constant $z_1$ determines the position, on the $z-$axis, of the black hole. When only one source is present this constant can be gauged away by the translation invariance property of the solution, along the $z$-axis, therefore is often omitted. However in our limiting procedure it could be useful, because we would like to grow the black hole size avoiding that the whole event horizon runs to spatial infinity.    \\
Before taking the limit for large mass and electric charge parameters we rescale the three parameters of the solution ($m,q,z_1$) and the coordinate as follows
\beq
        m \to \mezzo (w_2-w_1)(1+2w_2 \d^2) \ , \hspace{1.5cm} q \to  (w_1-w_2)\sqrt{2w_2} \d \ , \hspace{1.5cm} z_1 \to \mezzo (w_2+w_1)(1-2w_2\d^2) \ , \ \ 
\eeq 
\beq
            \rho \to \rho (1-2w_2\d^2) \ \ , \hspace{1.5cm} z \to  z (1-2w_2\d^2) \ \ ,  \hspace{1.5cm}  t \to \frac{\sqrt{A}}{\d^2 \sqrt{2w_2}} t \ \ , \hspace{1.5cm} \varphi \to \frac{\sqrt{A}}{\d^2 \sqrt{2w_2}} \varphi \ \ . \ \
\eeq
Now taking the limit for $w_2 \to \infty$ and rescaling $\d \to \d \sqrt{A}$ we get precisely the solution ({\ref{g-backgr-D})-(\ref{A-backgr-D}) with $c=0$. The similarities with (\ref{g-backgr-D})-(\ref{A-backgr-D}) can be better appreciated by the change of coordinates
\beq
        \r = \frac{r \sqrt{(1-A^2r^2)(1-x^2)}}{(1+Arx)^2} \ \ , \hspace{1.5cm} z = z_1+\frac{r (Ar+x)}{(1+Arx)^2} \ .
\eeq
 In case one would like to recover also the $c\neq0$ case it is necessary to perform this limit starting from a Reissner-Nordstrom-NUT spacetime.\\
This background is free from conical singularities, but the metric, for $c=0$, seems to be not free of curvature singularity for $0<r<1/A$, as can be appreciated from the Kretschmann scalar
\beq
           R_{\m\n\s\l} R^{\m\n\s\l} = \frac{64A^4s^4(1+Arx)^{12}[3(1-A^2r^2)^2s^4+5(1+Arx)^4+6s^2(1-A^2r^2)(1+Arx)^2]}{[-(1-A^2r^2)s^2+(1+Arx)^2]^8} \ .
\eeq
On the other hand when $c\neq 0$ the denominator of the main scalar invariants can never be zero, as can be seen from the Ricci squared\footnote{We write the Ricci squared and not the Kretschmann only for brevity, anyway the possible divergent loci coincides for both scalar invariants.}
\beq
          R_{\m\n} R^{\m\n} =  \frac{64A^4s^4(1+Arx)^{16}}{\{c^2(1-A^2r^2)^2+[-(1-r^2A^2)s^2+(1+Arx)^2]^2\}^4} \ \ ,
\eeq
therefore, in the general case, the solution ({\ref{g-backgr-D})-(\ref{A-backgr-D}) is free of any singularities, nor curvature or conical. The only locus of the spacetime which might seem suspect of bringing curvature singularities, for $c\neq 0$, is for $x=-1 , r=1/A$, but a deeper scrutiny of various scalar invariants confirms it is not a problematique point. \\
Note that this metric has, for any value of $c$, neither NUT charge nor Misner string because the Ehlers transformation rotates mass or electric charge into gravitomagnetic mass. When we set $m=0$, $e=0$ and $p=0$ we are also shutting down the NUT charge. That is also why the Petrov type remains D\footnote{One might think it should be a Melvin-like electric universe in some accelerating coordinates, but contrary to Melvin this background is not able to help to remove the conical singularity of the charged accelerating black hole. Furthermore the zero acceleration limit remove also the electromagnetic field, which is not in line with the usual Melvin behaviour in accelerating coordinates, such as the Ernst metrics \cite{ernst-remove}. Finally note that accelerating Melvin universe has not curvature singularities and can be thought as a couple of RN black hole pushed at spatial infinity, while in our case we push at most one of the two the black holes asymptotically, as described in sections \ref{sec:back} and \ref{limit}.}.\\
In any case possibly this type D solution, where we switched off the seed physical quantities of the balck hole, may be not the more adequate spacetime to be considered the background of our solution, because after the Harrison  transformation the characteristic quantities of the black hole change, as can be seen in appendix \ref{app:harrison}, and a change of coordinates is necessary. If we eliminate the black hole after the charging transformation and in the new parametrization, then we get a NUTty and charged Type I background, similarly to what happens in the pure  NUT case \cite{PD-NUTs}. \\
We write here a simple case without the $c$ parameter, because we already have it in the new parametrization. Indeed this spacetime can be obtained from the solution in section \ref{new-par-sol+e} by taking the limit for zero black hole electric charge $\bar{e}$ and zero black hole mass\footnote{Is not easy to compute the mass for accelerating black holes with conical singularities, therefore the mass parameter $\bar{m}$ is surely related with the black hole mass in the non accelerating case. We are aware that a different parametrisation might give different results for the type I background spacetime, this point deserves further investigation.} parameter $\bar{m}$:
\beq \label{background-I-inizio}
     ds^2 = -f(\bar{r},x) dt^2 + \frac{1}{f(\bar{r},x)} \left[ e^{2\gamma(\bar{r},x)}  \left( \frac{{d \bar{r}}^2}{\Delta_r(\bar{r})} + \frac{{d x}^2}{\Delta_x(x)} \right) + \rho^2(\bar{r},x) d\varphi^2 \right] \ ,
\eeq
where
\bea
f(\bar{r},x)      &:=& \frac{4 \Om^2\D_r}{\bar{r}^4[2x+(e+\bar{r})xA]^2}  \ \    \hspace{2.2cm}    \Omega(\bar{r},x) \ := \ 1 + A (e+\bar{r}) x \ , \nn \\
\g(\bar{r},x)     &:=& \mezzo \log \left( \frac{\D_r}{\Om^4} \right)  \ \ ,   \hspace{3.4cm}   \Delta_r(\bar{r}) \ :=\  \bar{r}^2[1-(e+\bar{r})^2A^2] \ ,\nn \\
\rho(\bar{r},x)   &:=& \frac{\sqrt{\D_r} \sqrt{\D_x}}{\Om^2} \ \ ,          \label{background-I-fine}  \hspace{3.7cm}  \Delta_x(x) \ := \ (1-x^2) (1+eAx)  \ . 
\eea
The electromagnetic potential supporting the this metric is
\beq
           A_\m = \left\{ - \frac{2[1+2exA+(e+\bar{r})(ex^2-\bar{r})A^2]}{2 \bar{r} x + \bar{r}(e+\bar{r})(1+x^2)A} , 0,0,0 \right\} \ .
\eeq
The limit is quite straightforward, the only care we used, to assure a convergent result, was to rescale the coordinates before the limiting procedure: $t \to (e\a/\bar{m})\ t$ , $\varphi \to \varphi \bar{m}/(e\a)$. \\
This solution is diffeomorphic to the extremal case ($m=e$) of (\ref{g-backgr-D}), for $c=0$ and $s=-1$, up to a radial shift. Because of the fact that $s \neq 0$ the standard Rindler background cannot be retrieved easily.\\
When or $e$ or $A$ goes to zero the metric became algebraically special, of type D. When the accelerating\footnote{After the reparameterisation and the limit of zero $\bar{m}$ , $\bar{e}$ the constant $A$ may not keep the usual acceleration interpretation.} parameter $A \to 0$ a cylindrical electric monopole is obtained:
\beq
         ds^2 = - \frac{\bar{q}^2}{\r^2} dt^2 + \frac{\bar{\r}^2}{\bar{q}^2} d\bar{\r}^2 + \bar{\r}^2 (d\bar{x}^2+d\bar{y}^2) ,
\eeq 
\beq
         A_\m = \left( - \frac{\bar{q}}{\bar{\r}} , 0 , 0, 0 \right)  \ .
\eeq
The following change of coordinates has been used to put the metric, after the $A\to0$ limit, in the above form:
\beq
             \bar{r} = \sqrt{\bar{q}^2 \bar{x}^2+\bar{q}^2 \bar{y}^2+\bar{\rho}^2/\bar{q}^2} \ ,  \qquad \varphi = \arctan \left( \frac{\bar{y}}{\bar{x}} \right) \ , \qquad x = \frac{\bar{\rho}}{\bar{q} \sqrt{\bar{q}^2 \bar{x}^2+\bar{q}^2 \bar{y}^2+\bar{\r}^2/\bar{q}^2}} \ .
\eeq
Hence the electric field of the background can be interpreted as a monopole charge and it is not a surprise that the metric (\ref{background-I-inizio})-(\ref{background-I-fine}) displays curvature singularities. \\
We expect the more general case, when $c\neq 0$, to be more regular, in particular to be void of curvature singularities, as happens for the type D background above.\\

\subsection{Misner string?}

Let's come back to the full solution of eq. (\ref{lwp-rx-new-charged-nut})-(\ref{lwp-rx-new-charged-nut-fine}). The Ehlers transformation brings into the spacetime the NUT charge and hence the Misner string, a rotating delta-like energy-density distribution on the azimuthal axis, which is not continuous because the $\omega(r,x)$ function has a jump passing from the sector of the $z$-axis characterised by $x\to1$ to $x\to-1$. We can quantify this discontinuity as
\beq
           \D\om =  \lim_{x \to 1} \bar{\omega} (r,x) - \lim_{x \to -1} \bar{\omega} (r,x) = 8 [c (m+es)-ps(1-s^2)]
\eeq
Spacetime without a Misner string experiences a zero $\D\om$. For the dyonic Reissner-Nordstrom-NUT black hole in an accelerating and  nutty background this value vanishes for
\beq
       c = \frac{ps (1-s^2)}{m+es} \ .
\eeq
Moreover the value of the rotating function $\om(r,x)$ can be made zero on the axis by setting
\beq
           \om_0 = - \frac{2ps(1-s^2)}{A(m+es)}
\eeq
Therefore the gravitomagnetic mass introduced by the coupling between the Harrison transformation and the magnetic charge (encoded into the term proportional to $ps$ in $\D\om$) can be fine tuned with the NUT contribution brought by the Ehlers parameter $c$. Indeed without the Ehlers map, as can be seen in the appendix \ref{app:dyonic-acc-RN-I}, the only way to regularise the metric is to switch off also the magnetic charge $p$. \\

\section{Summary, Discussion and Conclusions}

In this article we have explored the behaviour of the Harrison transformation on accelerating black holes. We have discovered that the Lie symmetry map acts on an accelerating black hole seed adding electromagnetic charge to the initial solution in a non-trivial way. In fact it does not generate the standard charged C-metric from the neutral one, but an {\it exotic} C-metric which is type-I instead of type-D, as the standard one. Both charged C-metric describe accelerating and charged black holes, but the type-I is more general, it has an extra feature: also the accelerating background can carry its own electromagnetic charge, which is independent with respect to the black hole one, as discussed in section \ref{sec:generation}. The two charges may interact and combine, for instance to discharge the black hole, transforming a seed accelerating Reissner-Nordstrom solution (of type D) in an exotic accelerating Schwarzschild (Type I), but the two electric charges cannot completely annihilate everywhere because of their physical differences: their distribution within the spacetime and their relative independence. While the action of the Harrison transformation on C-metrics can cause curvature singularities, these are absent for the combined Ehlers-Harrison transformation, because of the presence also of the NUT parameter.  \\
This behaviour resembles much what is described in \cite{PD-NUTs} about the accelerating NUT spacetime and the Ehlers transformation. Also in that case the NUT charge can be added to accelerating black hole and can combine to elide the Misner string of the seed, if present, but never remaining with a type-D metric.\\
These findings about the charging transformation strengthen the interpretation given in \cite{PD-NUTs} about the action of these non-trivial Lie-point symmetries and the nature of accelerating black holes and C-metrics, that we summarise here below: Accelerating black holes can be viewed as limits of binary system where one of the collapsed sources grows indefinitely while maintaining fixed the distance between the horizons\cite{bubble}, \cite{PD-NUTs}. Similarly one may think of zooming near the horizon of a big black hole which has a much smaller companion close by. In both interpretations the big black hole in this limit becomes an accelerating horizon. On the other hand the Harrison transformation acting on a neutral binary system equally adds electromagnetic charge to both the sources, for instance it can generate a charged binary system (such as the Majumdar-Papapetrou) from the neutral Bach-Weyl pair \cite{many-rotating}. In this charged setting the above limit where one element of the black hole couple grows indefinitely, as proven in section \ref{limit}, represents the exotic charged accelerating black hole we had described here. The fact that also the big black hole is endowed with electric charge causes the background to have non-zero electric field. According to this interpretation the standard (type-D) charged C-metric can be thought as the limit of a binary system where only the small black hole carries charges (not only electromagnetic but also NUT charges, as seen in \cite{PD-NUTs}) but the big black hole is neutral. This is the reason why these exotic accelerating black holes metrics can have double independent charges. One of these charges is reminiscent of the black hole charge that has been infinitely enlarged. In this sense is not surprising to find for this type-I charged and accelerating black holes that extremal configurations can be void of conical singularities, such as the Majumdar-Papapetrou system.\\
These results open to the possibility of building even more general accelerating black holes in general relativity and beyond\footnote{In fact the solution generating techniques here used can be extended to scalar tensor theories, such as Brans-Dicke or conformally coupled scalar fields \cite{marcoa-embedding},\cite{marcoa-stationary}. Some works on accelerating NUTty black holes with a conformally coupled scalar field have been recently obtained using these generating techniques \cite{adolfo}.}. In principle it is quite direct to apply the Harrison transformation to a charged and rotating C-metric seed to obtain an exotic accelerating type-I Kerr-Newman black hole, where the accelerating background is endowed with electromagnetic field or even with NUT charge using the combined transformation of section \ref{sec:generation-NUT}. While the result in terms of the Ernst potential is straightforward, as can seen in appendix \ref{app:full-PD}, the metric form of the solution could be quite involuted, in particular in a convenient parametrisation. Another interesting possibility is to add other kinds of charges to these accelerating backgrounds, for instance angular momentum, works in this direction are in progress. \\

\section*{Acknowledgements}
{\small We thank Giovanni Boldi, Silke Klemm and Adriano Viganò for interesting discussions on the subject and Roberto Emparan for stimulating comments. 
A Mathematica notebook containing the main solutions presented in this article can be found in the arXiv source folder. 
}\\

\appendix

\section{Dyonic Reissener-Nordstrom in a charged Rindler background}
\label{app:dyonic-acc-RN-I}

If we leave  $p \neq 0$ in the solution generated on section \ref{sec:generation} we obtain the dyonic version of the charged Reissner-Nordstrom black hole in the Rindler background. The metric becomes\footnote{A Mathematica file containing this solution is available between  the sources of the arXiv files, for the readers' convenience.}
\beq \label{lwp-rx-new-complete}
        ds^2 = - \frac{f(r,x)}{|1 -2s\mathbf{\Phi} - s^2 \Er |^2} \big[dt - \bar{\omega}(r,x) d\varphi\big]^2  + \frac{|1 - 2s \mathbf{\Phi} - s^2 \Er |^2}{f(r,x)} \left[ e^{2\gamma(r,x)} \left( \frac{{d r}^2}{\Delta_r(r)} + \frac{{d x}^2}{\Delta_x(x)} \right) + \rho^2(r,x) d\varphi^2 \right]  ,
\eeq
with 
\beq
       \bar{\omega}(r,x) = \frac{4sp\{(s^2-1)x+2A[s^2(r-m(1-x^2))-rx^2] +A^2x[s^2(r-q^2(1-x^2)-r^2x^2]\}}{\Omega^2}  + \omega_0 \ .
\eeq
where $\om_0$ is an arbitrary constant that usually defines the angular speed of the asymptotic observer. \\
The electromagnetic potential, up to an additive constant, takes the form 
\bea \label{barAt_new-complete}
 \bar{A}_t &=&  \frac{e r^3 \Omega^4 + s^3(\Omega^2 q^2-\D_r)^2 + 3 s^2 e r \Omega^2 (\Omega^2 q^2 - \D_r) + s r^2 \Omega^2 (3q^2\Omega^2-\D_r)}{4ser^3 \Omega^4 + r^4\Omega^4 + s^4 (\D_r - q^2 \Omega^2)^2 +4s^3er\Omega^2(\Omega^2 q^2 -\D_r) + 2 s^2 r^2 \Omega^2 (3 q^2 \Omega^2 - \D_r)} \  ,  \\
 \bar{A}_\varphi &=& \bar{A}_{\varphi_0} - 2px -\bar{\omega} \left(\frac{3}{4s} +\bar{A}_t \right) \nn \ .
\eea

As computed in section \ref{sec:generation}, this solution generically belongs to the I class of the Petrov classification. \\
The metric is not diagonal, as the seed, because there is an interaction due to the Lorentz force, between the electrically charged Rindler background and the intrinsic magnetic charge of the black hole $p$. From a mathematical point of view this fact can be directly read in the $sp$ coupling in the rotating $\omega$ function. Indeed if $s$ or $p$ are switched off the metric become diagonal.\\

\subsection{Misner string and gravitomagnetic mass}
While the appearance of Dirac strings is an expected feature from a complex charging transformation, what is not completely expected is that the rotation introduced by the Harrison map is not regular. In fact the transformation (\ref{harrison}) in the presence of a magnetic field switches on also an axial singularity, which can be associated to the a discontinuity of the $\omega(r,x)$ function passing through the equatorial plane on the  azimuthal axis, as can be seen from the difference
\beq \label{Dw}
      \D \omega =  \lim_{x \to 1} \bar{\omega} (r,x) - \lim_{x \to -1} \bar{\omega} (r,x) = - 8 s p (1-s^2) \ .
\eeq
This discontinuity physically represents a rotating delta-like matter distribution on the $z$-axis, often called NUT charge. In case we would like to remove this defect in the spacetime, we might act with an extra Ehlers transformation to rotate the gravitomagnetic mass quantified in eq (\ref{Dw}) into the usual mass, as explained in \cite{enhanced}. Note also that the gravitomagnetic mass generated by the Harrison transformation does not depend on the presence of the acceleration, in fact this can happen also in asymptotically flat conditions. Hence would be desirable to improve the standard Harrison transformation to avoid this behaviour when the seed carries monopolar magnetic charge.\\

\subsection{Enhanced Harrison transformation}
\label{app:harrison}

In order to have a more precise charging transformation, that is a symmetry of the Ernst equations which just adds electromagnetic charges on a given seed, but without messing with other physical parameters of the solution, we might try to modify the standard Harrison transformation $(V)$, as defined in (\ref{harrison}). We would like to refine this transformation composing with other symmetries of the Ernst equations, because in this way, we are sure that the spacetime it generates satisfies, by construction, the Einstein-Maxwell field equations (\ref{field-eq-g})-(\ref{field-eq-A}). We focus here, for simplicity to relaxed asymptotically flat spacetimes, which means that the Ernst potential decay for large values of the radial coordinate to the Minkowski ones ($\Er=1, \ \mathbf{\Phi}=0$), as follows
\bea \label{complex-asym} 
       \Er &\sim & 1- \frac{2 \left(M - i B\right)}{r} + \frac{(z_*+2iJ) x + const}{r^2} + O \left(\frac{1}{r^3}\right) \quad ,  \\
    \label{complex-asym2}   \mathbf{\Phi} &\sim &  \frac{Q_e + i Q_m}{r} + \frac{(D_e+iD_m) x + const}{r^2} + O\left(\frac{1}{r^3}\right)     \quad ,
\eea
with $M, B, J, Q_e, Q_m, D_e, D_m$, respectively are associated with the physical quantities of the solution: mass, NUT, angular momentum, electric and magnetic charge, electric and magnetic dipole moments, on the other and $z_*$ is a constant related to the position of the origin of the coordinates $(r,x)$. Relaxed is referring to the fact that the metric can approach asymptotically a little more general spacetime with respect to the Minkowski one, i.e. Taub-NUT.\\
Other symmetries of the Ernst equations (\ref{ee-ernst-ch})-(\ref{em-ernst}) which can be useful for improving the Harrison map are
\bea \label{I}
      (I):    && \Er \longrightarrow \Er' = \l \l^* \Er  \qquad \ \quad \qquad \ ,  \qquad \mathbf{\Phi} \longrightarrow  \mathbf{\Phi}' = \l \mathbf{\Phi}  \ , \\
      (IV):   && \Er \longrightarrow \Er' = \Er - 2\b^*\mathbf{\Phi} - \b\b^* \ \ \ , \qquad \mathbf{\Phi} \longrightarrow  \mathbf{\Phi}' = \mathbf{\Phi} + \b    \ ,
\eea   
where in general $\l$ and $\b$ are complex scalars which parametrise the transformation, see \cite{stephani-big-book} or \cite{enhanced} for details. We can define an enhanced version of the Harrison transformation $(\bar{V})$ thanks to the following composition
\beq \label{barV}
         (\bar{V}) \ :=  \   (IV) \circ (V) \circ (I)  ,
\eeq
with 
\bea
       \l   &=& \frac{-1+\sqrt{1+4| \a |^2}}{2 | \a |^2} \exp (i v) \ ,  \\
        \b  &=&  \a \ \frac{1-\sqrt{1+4| \a |^2}}{2 | \a |^2}  \ . 
\eea 
$(\bar{V})$ preserves the asymptotic form of the  Ernst complex potential  (\ref{complex-asym})-(\ref{complex-asym2}), in fact by applying  (\ref{barV}) we get
\bea 
       \bar{\Er} &\sim & 1- \frac{2 \left(\bar{M} - i \bar{B}\right)}{r} + \frac{(\bar{z}_*+2i\bar{J}) x + \bar{const}}{r^2} + O \left(\frac{1}{r^3}\right) \quad ,  \\
    \mathbf{\bar{\Phi}} &\sim &  \frac{\bar{Q}_e + i \bar{Q}_m}{r} + \frac{(\bar{D}_e+i\bar{D}_m) x + \bar{const}}{r^2} + O\left(\frac{1}{r^3}\right)     \quad ,
\eea
where the new barred physical quantities are related to the old ones by the following transformations
\bea
      \bar{M} &=&    M \sqrt{1+4|\a|^2} - 2 [Q_e Re(\a) +  Q_m Im(\a)] \cos v - 2 [Q_e Im(\a) -  Q_m Re(\a)] \sin v \ , \ \\
      \bar{B} &=&    B \sqrt{1+4|\a|^2} - 2 [Q_e Im(\a) - Q_m Re(\a)] \cos v +2[Q_e Re(\a) + Q_m Im(\a)] \sin v \ , \ \label{barB} \\
      \bar{z}^* &=&  z^* \sqrt{1+4|\a|^2} + 4 [D_e Re(\a) + D_m \ Im(\a) ] \cos v + 4 [D_e Im(\a) - D_m Re(\a)] \sin v \ ,  \\\
      \bar{J} &=&    J \sqrt{1+4|\a|^2} + 2 [D_e Im(\a) -  D_m \ Re(\a)] \cos v + 2 [D_e Re(\a) + D_m Im(\a)] \sin v  \ , \ \\
      \bar{Q}_e &=&  (Q_e \cos v  - Q_m \sin v) \sqrt{1+4|\a|^2} -2M Re(\a) -2 B \ Im(\a) \ , \ \\
      \bar{Q}_m &=&    (Q_m \cos v +Q_e \sin v) \sqrt{1+4|\a|^2} - 2M Im(\a) + 2 B \ Re(\a) \ , \ \label{barQm} \\
      \bar{D}_e &=&  (D_e \cos v -D_m \sin v) \sqrt{1+4|\a|^2} + z^* Re(\a) -2J\ Im(\a) \ , \ \\
      \bar{D}_m &=&  (D_m 'cos v +D_e \sin v) \sqrt{1+4|\a|^2} +z^* Im(\a)   + 2 J Re(\a) \ , 
\eea
This map between new and old quantities allows one to build the Harrison transformation with the desiderated features. For instance if we want not to add additional NUT charge to the generated solution it is sufficient, by inspecting (\ref{barB}), to require an extra constraint on the $v$ or the $\a$ parameters:
\beq
        Im(\a) = Re(\a) \ \frac{Q_m \cos v + Q_e \sin v}{Q_e \cos v -Q_m \sin v} \ .
\eeq
Or in  case we would like to erase the Dirac string of a solution charged by $(\bar{V})$ starting from a seed free of Misner strings ($B=0$) then we may require, from (\ref{barQm}), that
\beq
            Im(\a) = \pm \frac{\sqrt{1+4Re(\a)^2} \ (Q_e \sin v + Q_m \cos v)}{2 \sqrt{(M-Q_m\cos v -Q_e \sin v)(M+Q_m\cos v + Q_e \sin v)}} \ .
\eeq  \\

\section{Plebanski-Demianski in NUTty and charged Rindler background}
\label{app:full-PD}

For the sake of generality we can consider here also the most general solution of these accelerating type I black holes with a charged and NUTty Rindler background. We represent it in detail in terms of the Ernst potentials, since the metric expression is quite lengthy, however also the metric representation is reported below. As seed we consider the Plebanski-Demianski solution, which includes black holes with dyonic electromagnetic charges, NUT charges and angular momentum. This seed can be written as in eq (\ref{lwp-rx}), with
\bea
f(r,x)&:=& \frac{\hat{\om}^2 \D_x -\D_r}{\hat{\om} \Om^2  \mathcal{R}^2 }  \ \ , \\
\om(r,x)&:=& \frac{\hat{\om}(r^2 \D_x + x^2 \D_r)}{\D_r-\hat{\om}^2\D_x} \ \ , \\
\g(r,x)&:=& \mezzo \log \left( \frac{\D_r-\hat{\om}^2\D_r}{\Om^4} \right) \label{gamma-pd} \ \ , \\
\rho(r,x)&:=& \frac{\sqrt{\D_r} \sqrt{\D_x}}{\hat{\om} \Om^2} \ \ , \label{rho-pd} \\
\Delta_r(r)&:=& -\hat{\om} (e^2 + p^2 + k \hat{\om}^2) + 2 m \hat{\om} r - \e \hat{\om} r^2 + 2 \hat{n} \a r^3 + k \a^2 \hat{\om} r^4   \ \ ,   \\
\Delta_x(x)&:=& -k \hat{\om} - 2 \hat{n} x + \e \hat{\om} x^2 - 2 m \a \hat{\om} x^3 + \a^2\hat{\om}(e^2 + p^2 + k \hat{\om}^2) x^4  \ \ , \label{Dx-pd} \\
\Omega (r,x) &:=& 1 - \a r x \ \ , \\
\mathcal{R}(r,x)&:=& \sqrt{r^2+\hat{\om}^2 x^2} \ \ ,
\eea
and the non null components of the electromagnetic vector potential are
\bea \label{PD-A}
A_t(r,x) &:=& - \frac{e r + \hat{\om} p x}{\mathcal{R}^2} \ \ , \\
A_\varphi(r,x) &:=&  \frac{e \hat{\om} r x^2 -  p x r^2}{\mathcal{R}^2}\label{PD-A2} \ \ .
\eea
From the definitions of the Ernst potentials (\ref{def-Phi-Er})-(\ref{h-e}) stem $h, \tilde{A}_\varphi$ and the seed Ernst complex fields as follows
\bea
h(r,x) &=& \frac{2 \big\{n r + \hat{\om} \big[ m x -kr^2\a-x^2 \a (e^2+p^2+k\hat{\om}) \big] \big\}  }{\Om \mathcal{R}^2} \  \  , \\
\tilde{A}_\varphi (r,x) &=& \frac{e x \hat{\om} - p r}{\mathcal{R}^2} \  \  , \\
\Er (r,x) &=& \frac{r \hat{\om} \D_x + i\big\{ \Om^2 \hat{\om} \big[ -ikr\hat{\om} + x (e^2+p^2+k\hat{\om}^2) \big] +x\D_r\big\}}{rx \Om^2 \hat{\om}(-ir+\hat{\om}x) }   \label{PD-Phi-seed}\  \  ,  \\
\mathbf{\Phi} (r,x) &=& - \frac{e + ip}{r+i\hat{\om}x}  \  \  .
\eea
Using the combined Harrison-Ehlers transformation (\ref{commute}), we can generate a new pair of complex Ernst potentials $(\bar{\Er},\bar{\mathbf{\Phi}})$ as follows
\beq  \label{Ebarfull}
\bar{\Er}(r,x) =   \frac{-ir\hat{\om}\D_x+x\D_r+\hat{\om}[-ikr\hat{\om}+x)(q^2+k\hat{\om}^2)]\Om^2}{(c+is^2)(r\hat{\om}\D_x+ix\D_r)-\hat{\om}[r^2x+2(ip+e)rsx + ir(x^2+ikc-ks^2)\hat{\om}+(s^2-icx)(q^2+k\hat{\om}^2)]\Om^2} ,
\eeq
\beq \label{Phisbarfull}
\bar{\mathbf{\Phi}}(r,x) =  \frac{-irs\hat{\om}\D_x+sx\D_r+\hat{\om}[(ip+e)rx+(q^2+k\hat{\om}^2)sx-ikrs\hat{\om}]\Om^2}{(c+is^2)(r\hat{\om}\D_x+ix\D_r)-\hat{\om}[r^2x+2(ip+e)rsx + ir(x^2+ikc-ks^2)\hat{\om}+(s^2-icx)(q^2+k\hat{\om}^2)]\Om^2} .
\eeq
These complex functions represent the new solution, inequivalent with respect to the seed, of the Einstein-Maxwell theory. It includes basically all the accelerating type I black holes described in this article and more. In particular these solutions can describe black holes of the Kerr-Newman family.  \\
In case one wants to write the solution in terms of the metric fields and electromagnetic potential it is sufficient to exploit again the definitions (\ref{def-Phi-Er})-(\ref{h-e}). The new function $f(r,x)$ can be easily written as
\beq \label{f-bar}
\bar{f} = \frac{f}{1 + (i c -s^2) \Er  - 2 s  \mathbf{\Phi}} \ \ ,
\eeq
while the transformed rotating function,  $\bar{\omega}(r,x)$ and the magnetic part of the gauge potential $\bar{A}_\varphi$ are more involved. However for completeness we explicitly write 
\bea \label{w-bar}
      \bar{\omega}(r,x) &=& \frac{1}{r^2x^2 \hat{\om} (\hat{\om}^2 \D_x -\D_r)(1-\Om)\Om^4 } \bigg\{ 
    c^2 (\Om-1)\Big[r^2\D_r \big( \D_x+k\hat{\om}\Om^2 \big)^2 +x^2 \D_x \big(\D_r+\hat{\om} (q^2+k \hat{\om}^2) \Om^2 \big)^2 \Big]  \nn \\  
        &-& 2 c r x \Om^2 \bigg[r^2 \D_r (\Om-1)(\D_x+k\hat{\om}\Om^2) + 2r \bigg( mx^2\hat{\om} (\D_r-\hat{\om}^2\D_x)\Om^2 +es\D_r(\Om-1)(\D_x+k\hat{\om}\Om^2) \bigg)   \nn \\
        &+& x\bigg( -x\D_r^2\Om + \hat{\om}^2(q^2+k\hat{\om}^2)\D_x\Om^2 (x\hat{\om}^2-2ps+2ps\Om) + \hat{\om} \D_r \Big(x(q^2+k\hat{\om}^2) (\Om-2)\Om^2 \nn \\
        &+&  \D_x \big( 2(ps+x\hat{\om})\Om-2ps-x\hat{\om} \big) \Big) \bigg) \bigg] + (\Om-1)  \bigg[x^2\hat{\om}^2 \Big(r^4+4er^3s+6q^2r^2s^2+4ers^3(q^2+k\hat{\om}^2) \Big)\D_x\Om^4 \nn \nn \\
        &+& s^4x^2\D_r^2\D_x  + \D_r \bigg(r^4s^4\D_x^2+2s^3 \Big( kr^2s\hat{\om} -2psr^2x + (2er+q^2 s)x^2\hat{\om} +ksx^2\hat{\om}^3 \Big) \D_x\Om^2 \nn \\ 
        &+& r^2\om (4psx^3-4kps^3x + k^2s^4\hat{\om} + x^4\hat{\om})\Om^4 \bigg) \bigg] \bigg\} \ .
\eea
The above expressions of the rotational function (\ref{w-bar}) with (\ref{f-bar}) and (\ref{gamma-pd})-(\ref{Dx-pd}) completely determine the new generated metric, which can be written, similarly to (\ref{lwp-rx}), as
\beq \label{lwp-rx-bar}
        d\bar{s}^2 = -\bar{f}(r,x) \left[ dt - \bar{\omega}(r,x) d\varphi \right]^2 + \frac{1}{\bar{f}(r,x)} \left[ e^{2\gamma(r,x)}  \left( \frac{{d r}^2}{\Delta_r(r)} + \frac{{d x}^2}{\Delta_x(x)} \right) + \rho^2(r,x) d\varphi^2 \right] \ .
\eeq

Actually the electromagnetic field supporting the metric can be conveniently derived from the electromagnetic Ernst potential, without the need of integrating explicitly $\bar{A}_\varphi$ from (\ref{A-tilde-e}). In fact the Faraday tensor for axisymmetric and stationary spacetime in the form of (\ref{lwp-rx-bar}) can be written as 

\begin{equation}
 \bar{F}_{\m\n} =
\begin{pmatrix}
0 & \p_t \bar{A}_t & \p_x \bar{A}_t & 0 \\
-\p_t\bar{A}_t & 0 & 0 &  \sqrt{\frac{\D_x}{\D_r}} \frac{\r \p_x \bar{\tilde{A}}_{\varphi}}{\bar{f}} + \bar{\om} \p_r \bar{A}_t \\
-\p_x \bar{A}_t & 0 & 0 & - \sqrt{\frac{\D_r}{\D_x}} \frac{\r \p_r \bar{\tilde{A}}_{\varphi}}{\bar{f}} + \bar{\om} \p_x \bar{A}_t \\
0 & - \sqrt{\frac{\D_x}{\D_r}} \frac{\r \p_x \bar{\tilde{A}}_{\varphi}}{\bar{f}} - \bar{\om} \p_r \bar{A}_t &  \sqrt{\frac{\D_r}{\D_x}} \frac{\r \p_r \bar{\tilde{A}}_{\varphi}}{\bar{f}} - \bar{\om} \p_x \bar{A}_t & 0 \\
\end{pmatrix} \ , \\
\end{equation}
where, from definition (\ref{def-Phi-Er}), $\bar{A}_t (r,x) = Re(\bar{\mathbf{\Phi}})$ and $\bar{\tilde{A}}_t (r,x) = Im(\bar{\mathbf{\Phi}})$. In case $\rho$ can be written similarly to (\ref{rho-pd}) the Faraday tensor can be further simplified. This solution is available, as a Mathematica notebook, between the arXiv files.\\
We observe that the metric built in this section represents the most general Plebanski-Demianski metric that can be built by the composition of the Ehlers and the Harrison transformation (\ref{commute}), even though we have considered just a real parameter labelling the Harrison transformation. As said in the above sections, this restriction, just eases the computation and thus simplifies the resulting metric without compromising the generality of the spacetime. The main reason is that, when an Ehlers-Harrison transformation is applied to a dyonic metric the electromagnetic Ernst potential is already complex therefore the imaginary part of the Ehlers-Harrison parameter can be reabsorbed just in a relabelling of the electromagnetic parameter of the seed Ernst potential. Alternatively it can be understood as the phase space degrees of freedom are already completely saturated by a real-parameter Ehlers-Harrison transformation for dyonic seeds. We can easily prove this fact considering a real parameters Ehlers-Harrison transformation, i.e. (\ref{commute}) with $\a=s$, with $f$ and $h$, for a dyonic electromagnetic seed field as described in (\ref{PD-Phi-seed}). The gravitational Ernst potential takes the form
\beq \label{Er-cosa}
     \bar{\Er} =  \frac{-i\big[e^2+p^2-(r^2+x^2 \hat{\om}^2)(f-ih)\big]}{c(e^2+p^2)+i\big[r^2+2(e+ip)rs+(e^2+p^2)s^2+2(p-ie)sx\hat{\om}+x^2\hat{\om}^2\big] - (c+is^2)(r^2+x^2\hat{\om}^2)(f+ih)}  \ .  
\eeq
Then we relabel\footnote{This relabelling of electromagnetic charges can be alternatively interpreted as an unitary electromagnetic rotation transformation (I) as in (\ref{I}) with $\l=exp(ia)$ ($a\in \mathbb{R}$), which in general is always an identity transformation for the gravitational Ernst potential, and a rotation of the Ernst electromagnetic potential. But in case of dyonic electromagnetic field, such as the one we are working with, the (I) transformation trivializes into the identity operator also for $\mathbf{\Phi}$ because it can be reabsorbed in a redefinition of the electromagnetic parameters.} the electric and magnetic charge such as 
\beq \label{ecose}
         \left\{\begin{matrix}
  &e \longrightarrow \ +e \cos a + p \sin a   \\
 &  p \longrightarrow \ - e \sin a + p \cos a & \\
\end{matrix}\right. \quad , \hspace{1.4cm}  \textrm{with} \hspace{0,9 cm} a = \arccos \left[ \frac{Re(\tilde{\a})}{ | \tilde{\a} | } \right]
\eeq
and we rename the real parameter $\a=s$ \footnote{The symbol $\tilde{\a}$ is used to  point to a full complex quantity and to distinguish it from the above real choice $\a=s$.} of the Harrison transformation 
\beq  \label{s-alpha}
      s \longrightarrow | \tilde{\a} | \ \ .
\eeq 
Hence the gravitational Ernst potential (\ref{Er-cosa}) becomes  
\beq
     \bar{\Er}(r,x) =    \frac{\Er}{1 + i c \Er - \tilde{\a} \tilde{\a}^* \Er -2\tilde{\a}^*  \mathbf{\Phi}}                \ ,  
\eeq
which exactly corresponds to the Ehlers-Harrison transformed gravitational Ernst potential with $\tilde{\a} \in \mathbb{C}$, as in (\ref{commute}). Similarly also the electromagnetic Ernst potential of a dyonic seed under a real parameter ($s$) Harrison transformation coincides with the one of a complex parameter ($\tilde{\a}$) Harrison map, up to a unitary electromagnetic duality rotation (\ref{I}), with $\l=\exp(ib)$. In fact acting with the $(I)$ transformation on the seed electromagnetic Ernst potential we get
\beq
  \bar{\mathbf{\Phi}} =   \frac{\exp(ib)\big[-(p-ie)(ir+ps+ies+x\hat{\om}) + s(r^2+x^2\hat{\om}^2) (f+ih) \big]}{-ic(e^2+p^2)+r^2+2(ip+e)rs+(e^2+p^2)s^2+2(p-ie)sx\hat{\om} + x^2\hat{\om}^2 + i(c+is^2)(r^2+x^2\hat{\om}^2) (f+ih)}        \    .
\eeq
Then considering the redefinitions in (\ref{ecose})-(\ref{s-alpha}) and 
\beq
                 b = \arctan \left[ \frac{Im(\tilde{\a})}{Re(\tilde{\a})} \right] \ \ ,
\eeq
we precisely recover the full complex Ehlers-Harrison transformation for the Ernst electromagnetic potential, as in (\ref{commute}):
\beq
       \bar{\mathbf{\Phi}}(r,x) = \displaystyle  \frac{ \mathbf{\Phi} + \tilde{\a} \Er }{1 + i c \Er - \tilde{\a} \tilde{\a}^* \Er -2\tilde{\a}^*  \mathbf{\Phi}} \ . 
\eeq
Note that the extra unitary $(I)$ transformation leaves $\bar{\Er}$ invariant. \\

Therefore we have proven that for a dyonic seed, such as the one we are considering in this section, the Ehlers-Harrison transformation (\ref{commute}) with real parameters is sufficient to generate the most general metric. Actually a complex $\tilde{\a}$ parameter does not provide an extended physical spacetime, but eventually only a more involved solution with fictitious parameters, which should be better to be reabsorbed to pursue a physical interpretation of the novel metric. The same observation, about the unnecessary complex parametrisation, when dealing with dyonic seeds, holds when $c=0$, i.e. for the standard Harrison symmetry (\ref{harrison}).\\


\begin{thebibliography}{99}


\bibitem{mann-stelea-chng}
B.~Chng, R.~B.~Mann and C.~Steequivalentlea,
{\it ``Accelerating Taub-NUT and Eguchi-Hanson solitons in four dimensions''},
 \href{https://doi.org/10.1103/PhysRevD.74.084031}{Phys. Rev. D \textbf{74} (2006), 084031} ;
\href{https://arxiv.org/pdf/gr-qc/0608092.pdf}{\tt [arXiv:gr-qc/0608092]}

\bibitem{Podolsky-nut}
J.~Podolsky and A.~Vratny,
{\it ``Accelerating NUT black holes''}, 
 \href{https://doi.org/10.1103/PhysRevD.102.084024}{Phys. Rev. D \textbf{102} (2020) no.8, 084024}; 
\href{https://arxiv.org/pdf/2007.09169.pdf}{\tt [arXiv:2007.09169 [gr-qc]]}. 


\bibitem{Plebanski-Demianski}
  J.~F.~Plebanski and M.~Demianski,
  {\it ``Rotating, charged, and uniformly accelerating mass in general relativity''},
  \href{https://doi.org/10.1016/0003-4916(76)90240-2}{Annals Phys. \textbf{98} (1976), 98-127}.

\bibitem{tesi-giova}
  G.~Boldi,
    {\it ``Ehlers transformation and accelerating spacetimes with a gravomagnetic monopole''},
     \href{https://inspirehep.net/literature/2652409}{Università degli Studi di Milano (2022)}

\bibitem{PD-NUTs}
  M.~Astorino and G. Boldi,
     {\it ``Plebanski-Demianski goes NUTs (to remove the Misner string)''},
     \href{https://doi.org/10.1007/JHEP08(2023)085}{JHEP \textbf{08} (2023), 085};
 \href{https://arxiv.org/pdf/2305.03744.pdf}{\tt [arXiv:2305.03744 [gr-qc]]}.
  
  
\bibitem{ernst-remove}
  F.~J.~Ernst,
  {\it ``Removal of the nodal singularity of the C-metric''}, 
   \href{http://scitation.aip.org/content/aip/journal/jmp/17/4/10.1063/1.522935}{J. Math. Phys. {\bf 17}, 515 (1976)}.


\bibitem{ernst-generalized-c} 
  F.~J.~Ernst,
  {\it ``Generalized C-metric''},
   \href{https://doi.org/10.1063/1.523896}{J.\ Math.\ Phys.\  {\bf 19}, 1986-1987 (1978).}


\bibitem{multipolar-acc}
M.~Astorino and A.~Vigan\`o,
{\it``Many accelerating distorted black holes''},
\href{https://doi.org/10.1140/epjc/s10052-021-09693-6}{Eur. Phys. J. C \textbf{81} (2021) no.10, 891};
\href{https://arxiv.org/pdf/2106.02058.pdf}{\tt [2106.02058 [gr-qc]]}.

\bibitem{ernst2}
  F.~J.~Ernst,
  {\it ``New Formulation of the Axially Symmetric Gravitational Field Problem. II''},
  \href{https://doi.org/10.1103/PhysRev.168.1415}{\tt Phys.\ Rev.\  {\bf 168} (1968) 1415.}

\bibitem{ernst-hauser}
I.~Hauser and F.~J.~Ernst,
{\it ``Proof of a generalized Geroch conjecture for the hyperbolic Ernst equation''},
\href{https://doi.org/10.1023/A:1002701301339}{Gen. Rel. Grav. \textbf{33} (2001), 195-293}; 
\href{https://arxiv.org/pdf/gr-qc/0002049.pdf}{\tt [arXiv:gr-qc/0002049 [gr-qc]]}.

\bibitem{many-rotating}
M.~Astorino and A.~Vigan\`o,
{\it ``Charged and rotating multi-black holes in an external gravitational field''},
\href{https://doi.org/10.1140/epjc/s10052-022-10787-y}{Eur. Phys. J. C \textbf{82} (2022) no.9, 829}, 
\href{https://arxiv.org/abs/2105.02894}{\tt [arXiv:2105.02894 [gr-qc]]}.

\bibitem{reina-treves}
  A.~Reina and A. Treves
  {\it ``NUT-like generalization of axisymmetric gravitational fields''} ,  \href{https://doi.org/10.1063/1.522614}{Journal of Mathematical Physics 16, 834 (1975)}.

\bibitem{enhanced}
  M.~Astorino,
  {\it ``Enhanced Ehlers Transformation and the Majumdar-Papapetrou-NUT Spacetime''},
  \href{https://doi.org/10.1007/JHEP01(2020)123}{JHEP \textbf{01} (2020), 123} ; 
   \href{https://arxiv.org/pdf/1906.08228}{\tt [arXiv:1906.08228 [gr-qc]]}.
    
\bibitem{ernst-magnetic} 
  F.~J.~Ernst,
  {\it ``Black holes in a magnetic universe''},
  \href{https://doi.org/10.1063/1.522781}{J.\ Math.\ Phys.\  {\bf 17}, no. 1, 54 (1976).}

\bibitem{marcoa-pair}
  M.~Astorino,
  {\it ``Pair Creation of Rotating Black Holes''},
  \href{https://doi.org/10.1103/PhysRevD.89.044022}{Phys. Rev. D \textbf{89} (2014) no.4, 044022};
  \href{https://arxiv.org/pdf/1312.1723.pdf}{\tt [arXiv:1312.1723~[gr-qc]]}.

\bibitem{swirling}
M.~Astorino, R.~Martelli and A.~Vigan\`o,
{\it ``Black holes in a swirling universe''},
\href{https://doi.org/10.1103/PhysRevD.106.064014}{Phys. Rev. D \textbf{106} (2022) no.6, 064014} ;
\href{https://arxiv.org/pdf/2205.13548}{\tt [arXiv:2205.13548 [gr-qc]]}.

\bibitem{marcoa-thermo}
M.~Astorino,
{\it ``Thermodynamics of Regular Accelerating Black Holes''},
\href{https://doi.org/10.1103/PhysRevD.95.064007}{Phys. Rev. D \textbf{95} (2017) no.6, 064007}
\href{https://arxiv.org/pdf/1612.04387.pdf}{\tt [arXiv:1612.04387 [gr-qc]]}

\bibitem{marcoa-removal}
M.~Astorino,
{\it ``Removal of conical singularities from rotating C-metrics and dual CFT entropy''}
\href{https://doi.org/10.1007/JHEP10(2022)074}{JHEP \textbf{10} (2022), 074};
\href{https://arxiv.org/pdf/2207.14305}{\tt [arXiv:2207.14305 [gr-qc]]}.


\bibitem{stephani-big-book}
  H.~Stephani, D.~Kramer, M.~A.~H.~MacCallum, C.~Hoenselaers and E.~Herlt,
  {``Exact solutions of Einstein's field equations''}, \ 
  \href{https://doi.org/10.1017/CBO9780511535185}{\tt [doi:10.1017/CBO9780511535185]}

\bibitem{bubble}
M.~Astorino, R.~Emparan and A.~Vigan\`o,
{\it ``Bubbles of nothing in binary black holes and black rings, and viceversa''},
\href{https://doi.org/10.1007/JHEP07(2022)007}{JHEP \textbf{07} (2022), 007} ;
\href{https://arxiv.org/pdf/2204.09690.pdf}{\tt [arXiv:2204.09690 [hep-th]]}.

\bibitem{majumdar}
S.~D.~Majumdar,
{\it ``A class of exact solutions of Einstein's field equations''},
\href{https://doi.org/10.1103/PhysRev.72.390}{Phys. Rev. \textbf{72} (1947), 390-398}

\bibitem{belinski-book}
V. Belinski, E. Verdaguer, \href{https://doi.org/10.1017/CBO9780511535253}{\it ``Gravitational solitons''}, Cambridge, Cambridge Univ. Press, 2001.


\bibitem{alekseev-belinski-2RN}
G.~A.~Alekseev and V.~A.~Belinski,
{\it ``Superposition of fields of two Reissner - Nordstrom sources''},
\href{https://arxiv.org/abs/0710.2515}{\tt [arXiv:0710.2515 [gr-qc]]}.


\bibitem{manko-2007}
V.~S.~Manko,
{\it ``The Double-Reissner-Nordstrom solution and the interaction force between two spherically symmetric charged particles''},
Phys.\ Rev.\ D {\bf 76} (2007) 124032;\
\href{https://doi.org/10.1103/PhysRevD.76.124032}{\tt [arXiv:0710.2158 [gr-qc]]}.

\bibitem{compere-kerr-cft}
G.~Comp\`ere,
{\it ``The Kerr/CFT correspondence and its extensions''},
\href{https://doi.org/10.1007/s41114-017-0003-2}{Living Rev. Rel. \textbf{15} (2012), 11} ;
\href{https://arxiv.org/pdf/1203.3561}{\tt [arXiv:1203.3561 [hep-th]]}.

\bibitem{lucietti-kunduri}
H.~K.~Kunduri, J.~Lucietti and H.~S.~Reall,
{\it ``Near-horizon symmetries of extremal black holes''},
\href{https://doi.org/10.1088/0264-9381/24/16/012}{Class. Quant. Grav. \textbf{24} (2007), 4169-4190} ;
\href{https://arxiv.org/pdf/0705.4214}{\tt [arXiv:0705.4214 [hep-th]]}.

\bibitem{strominger-kerr-cft}
M.~Guica, T.~Hartman, W.~Song and A.~Strominger,
{\it ``The Kerr/CFT Correspondence''},
\href{https://doi.org/10.1103/PhysRevD.80.124008}{Phys. Rev. D \textbf{80} (2009), 124008} ;
\href{https://arxiv.org/pdf/0809.4266}{\tt [arXiv:0809.4266 [hep-th]]}.

\bibitem{strominger-duals}
T.~Hartman, K.~Murata, T.~Nishioka and A.~Strominger,
{\it ``CFT Duals for Extreme Black Holes''},
\href{https://doi.org/10.1088/1126-6708/2009/04/019}{JHEP \textbf{04} (2009), 019} ;
\href{https://arxiv.org/pdf/0811.4393}{\tt [arXiv:0811.4393 [hep-th]]}.

\bibitem{acc-cft}
M.~Astorino,
{\it ``CFT Duals for Accelerating Black Holes''}
\href{https://doi.org/10.1016/j.physletb.2016.07.019}{Phys. Lett. B \textbf{760} (2016), 393-405} ;
\href{https://arxiv.org/pdf/1605.06131}{\tt [arXiv:1605.06131 [hep-th]]}.


\bibitem{melvin-lambda}
M.~Astorino,
{\it ``Charging axisymmetric space-times with cosmological constant''},
\href{https://doi.org/10.1007/JHEP06(2012)086}{JHEP \textbf{06} (2012), 086} ;
\href{https://arxiv.org/pdf/1205.6998.pdf}{\tt [arXiv:1205.6998 [gr-qc]]}.


\bibitem{adolfo}
J. Barrientos and A.~Cisterna,
{\it ``Ehlers Transformations as a Tool for Constructing Accelerating NUT Black Holes''},
\href{https://doi.org/10.1103/PhysRevD.108.024059}{Phys. Rev. D \textbf{108} (2023) no.2, 024059}; 
\href{https://arxiv.org/pdf/2305.03765.pdf}{\tt[arXiv:2305.03765 [gr-qc]]}.

\bibitem{marcoa-embedding}
M.~Astorino,
{\it ``Embedding hairy black holes in a magnetic universe''},
\href{https://doi.org/10.1103/PhysRevD.87.084029}{Phys. Rev. D \textbf{87} (2013) no.8, 084029} ;
\href{https://arxiv.org/pdf/1301.6794}{\tt [arXiv:1301.6794 [gr-qc]]}.

\bibitem{marcoa-stationary}
M.~Astorino,
{\it ``Stationary axisymmetric spacetimes with a conformally coupled scalar field''},
\href{http://doi.org/10.1103/PhysRevD.91.064066}{Phys. Rev. D \textbf{91} (2015), 064066} ;
\href{https://arxiv.org/pdf/1412.3539}{\tt [arXiv:1412.3539 [gr-qc]]}.

\end{thebibliography}
\end{document}